\documentclass[12pt,preprint]{aastex}
\usepackage{lscape}
\begin{document}

\title{INFRARED ECHELLE SPECTROSCOPY OF PALOMAR~6 AND M71\footnotemark[1]}

\author{Jae-Woo Lee\altaffilmark{2}, Bruce W. Carney\altaffilmark{3}
\and Suchitra C. Balachandran\altaffilmark{4}}

\footnotetext[1]{Based on observations made with the Infrared Telescope
Facility which is operated by the University of Hawaii under contract to
the National Aeronautics and Space Administration.}
\altaffiltext{2}{Department of Astronomy \& Space Science,
Astrophysical Research Center for the Structure and
Evolution of the Cosmos,
Sejong University, 98 Gunja-Dong, Gwangjin-Gu, Seoul, 143-747;
jaewoo@arcsec.sejong.ac.kr}
\altaffiltext{3}{Department of Physics \& Astronomy,
University of North Carolina, Chapel Hill, NC 27599-3255;
bruce@physics.unc.edu}
\altaffiltext{4}{Department of Astronomy, University of Maryland,
College Park, MD 20742; suchitra@astro.umd.edu}

\begin{abstract}
We present high-resolution infrared echelle spectroscopy for
the globular clusters Palomar~6 and M71.
Our mean heliocentric radial velocity of Palomar~6 is
+180.6 $\pm$ 3.2 km sec$^{-1}$ and is 20 km sec$^{-1}$ lower
than that found by Minniti in 1995.
Contrary to the previous metallicity estimates using low-resolution
spectroscopy, our results show that Palomar~6 has an intermediate
metallicity with  [Fe/H] = $-$1.0 $\pm$ 0.1
and is slightly more metal-poor than M71.
Reasonable changes in the surface temperature or
the microturbulent velocity of the model atmospheres do not affect
[Fe/H] at more than $\pm$ 0.2~dex.
In spite of its high metallicity,
based on the spectrum of a single cluster member
the [Si/Fe] and [Ti/Fe] ratios  of Palomar~6 appear to be enhanced by
0.4~dex and 0.5~dex, respectively, suggesting that the Galactic inner halo
may have experienced a very rapid chemical enrichment history.
\end{abstract}

\keywords{Galaxy: bulge --- globular clusters: individual (Palomar~6, M71) ---
stars: abundances --- infrared: stars}

\section{INTRODUCTION}
The measurement of [$\alpha$/Fe] ratios of globular clusters
provides important information regarding relative ages.
The $\alpha$-elements (C, O, Ne, Mg, Si, S, Ar, Ca, and Ti)
are synthesized predominantly
by massive stars, and ejected into the interstellar medium
by Type II supernovae (SNe~II) explosions.
On the other hand, iron-peak elements
are produced by both SNe~II and Type Ia supernovae (SNe~Ia).
Enhanced values of [$\alpha$/Fe] ratios in globular clusters
indicate the domination of SNe~II nucleosynthesis, while
lower ratios indicate increasing contribution of SNe~Ia, which
are thought to appear 10$^9$ or more years later.
Therefore, the measurement of the behavior of [$\alpha$/Fe]
versus [Fe/H] tells us
how rapidly chemical enrichment proceeded in our Galaxy
(Wheeler, Sneden, \& Truran 1989).
The [$\alpha$/Fe] ratio is also important in deriving
accurate age from comparisons of color-magnitude diagrams (CMDs)
to model isochrones.

[$\alpha$/Fe] ratios also provide a means of diagnosing the initial mass
function (IMF) and star formation rate (SFR) parameters (McWilliam 1997).
An IMF skewed to high-mass stars would have a higher $\alpha$-element yield
while a high SFR would give an enhanced [$\alpha$/Fe] at a higher [Fe/H]
since the interstellar medium will reach higher [Fe/H] before
the onset of SNe~Ia contributions.
In the recent study of three metal-poor inner old halo globular clusters
(Zinn 1993, Da Costa \& Armandroff 1995),
NGC~6287, NGC~6293, and NGC~6541, Lee \& Carney (2002b) claimed that
[Si/Ti] ratios are anticorrelated with Galactocentric distance of
the ``old" halo globular clusters, in the sense that the silicon abundance
increases and the titanium abundance decreases with Galactocentric distances,
and they proposed that the masses of
SNe~II progenitors may have been responsible for this elemental
abundance gradient.
However, the impact of the skewed IMF of SNe~II progenitors
in metal-rich globular clusters' $\alpha$-element abundances appears
to be unclear as yet mainly due to the lack of study of metal-rich clusters.
Thus more chemical abundance studies of metal-rich globular clusters
near the Galactic center are necessary,
since the most metal-rich globular clusters are concentrated in
the Galaxy's central regions.

Palomar~6 is known to be a metal-rich globular cluster located 0.9 kpc from
the Galactic center. As Lee \& Carney (2002a) discussed, the metallicity
of Palomar~6 has been in controversy for two decades.
Malkan (1981) was the first to estimate the metallicity of Palomar~6,
[Fe/H] = $-$1.30, using a reddening-free metallicity index $Q_{IR}$,
ranking it a metal poor inner halo cluster.
Zinn (1985) derived  [Fe/H] = $-$0.74 for Palomar~6
by reanalyzing Malkan's photometry.
Ortolani et al.\ (1995) estimated [Fe/H] $\approx$ $-$0.4 based on the
($V$, $V-I$) - CMD.
The results of Minniti (1995b) and Bica et al.\ (1998) are worth noting.
Minniti (1995b) studied the metallicities of seven highly reddened
clusters near the Galactic center including Palomar~6.
He obtained low S/N ratio spectra of six red-giant branch (RGB) stars
with 2 \AA\ resolution covering
$\lambda\lambda$ 4700 - 5400 \AA. He suggested that the sum of
Mg $\lambda$ 5175 \AA, Fe $\lambda$ 5270 \AA, and
Fe $\lambda$ 5535 \AA\ lines (Mg + 2Fe) is the optimal indicator of
metallicity following Faber et al.\ (1985),
and Minniti derived [Fe/H] = +0.2 $\pm$ 0.3 for
Palomar~6 based on the location of the giants
in the (Mg + 2Fe) versus $(J-K)_0$ diagram (see his Figure~1).
More recently, Bica et al.\ (1998) undertook a metallicity study of
the globular clusters near the Galactic center with an expanded sample.
They measured the Ca~II triplet $\lambda\lambda$ 8498, 8542,
and 8662 \AA\ using intermediate resolution (5.4 \AA~pixel$^{-1}$)
integrated spectra and obtained [$Z/Z_{\odot}$] = $-$0.09
for Palomar~6.
It should be emphasized that the result of Bica et al.\ (1998) may have
a potential problem with field star contamination, since Palomar~6
lies in the Galactic center and the field star contamination is expected
to be very high (see, for example, Figure~8 of Lee \& Carney 2002a).
The abundance study of Galactic bulge giants showed that the mean
[Fe/H] = $-$0.25 (McWilliam \& Rich 1994), which is comparable to that of
Bica et al.\ (1998) found for stars in or near Palomar~6.
Also the metallicity estimates relying on low resolution spectra can give
erroneous results in the metal-rich RGB stars due to blending
with some of the notorious molecular bands, such as TiO
(see, for example, Plez 1998) and
CN (see, for example, McWilliam \& Rich 1994).
McWilliam \& Rich (1994) have demonstrated the effect of
the CN blending in the metal-rich bulge K giants in their Figure~7.
They claimed that the CN blending may cause a discrepancy in metallicity
by $\approx$ 0.3 dex between the results of McWilliam \& Rich (1994),
who employed high resolution spectroscopy, and those of Rich (1988),
who employed a similar method as Minniti (1995).
Recently, Stephens \& Frogel (2004) employed medium-resolution
($R$ = 1380 -- 1650) infrared (IR) $K$-band spectroscopy for
8 globular clusters near the Galactic center.
They measured Na~I doublet ($\lambda\lambda$ 21910 -- 21966\AA),
Ca~I triplet ($\lambda\lambda$ 22450 -- 22560\AA),
CO(2,0) band ($\lambda\lambda$ 22915 -- 23025\AA) and obtained
[Fe/H] = $-$0.52 $\pm$ 0.11 for Palomar~6.
Comparisons of high resolution spectra
of cool stars of Wallace \& Hinkle (1996) with the spectral regions
used by Stephens \& Frogel (2004) show that their spectral
regions are vulnerable to contamination from the CN blending
and their results may have problems similar to those of Rich (1988).
Therefore high resolution spectroscopy free from blending with
molecular bands is required.

In this paper, we present high-resolution IR spectroscopy
of three RGB stars in Palomar~6.
Due to its high interstellar reddening value,
$E(B-V)$ $\approx$ 1.25 -- 1.4 mag, IR spectroscopy is essential for
studying Palomar~6.

\section{OBSERVATIONS AND DATA REDUCTION}
Our observations were carried out at the 3.5m NASA Infrared Telescope
Facility using CSHELL, the facility cryogenic IR echelle
spectrograph (Tokunaga et al.\ 1990; Greene et al.\ 1993),
on July 13 -- 14 1997 for M71 and May 22 -- 25 1998 for Palomar~6.
CSHELL is a long-slit spectrograph which uses a 31.6 lines/mm, 63.5$\arcdeg$
echelle grating with narrow band circular variable filters that isolate
a single order. The detector for our observations was
a 256 $\times$ 256 Hughes SBRC InSb array, which provides a spectral coverage
of $\approx$ 750 km sec$^{-1}$ ($\approx$ 50 \AA).
All of our observations were obtained with a 0.5 arcsec slit, providing
a nominal resolution of 42,000, or 7 km sec$^{-1}$,
using a ``nodding" mode, which moves the telescope
between frames and stores each frame separately.
Because of the limited spectral coverage in a single grating setting,
it was necessary to choose particular  grating settings to cover
spectral features of special interest.
We chose two spectral regions  based on the IR Atlas of Arcturus
(Hinkle et al.\ 1995).
One setting is for the two Fe~I lines
(centered at $\approx$ $\lambda$ 22395 \AA)
and the other is for the two Si~I lines and a Ti~I line
(centered at $\approx$ $\lambda$ 21815 \AA).
It should be emphasized that our spectral regions are free of
molecular lines, such as CN or CO.
Both spectral regions contain the neutral scandium lines, however,
we do not derive the scandium abundance due to the lack of information
of the hyperfine splitting of the Sc~I lines.

The basic photometric data for our targets using star IDs of
Minniti (1995b) are given in Table~1 and
the finding chart is given in Figure~1.
Note that we adopt $JHK$ photometry by Lee \& Carney (2002a) in the Table.

We also obtained spectra for the $^{12}$CO 2-0 bandhead
for each target and for Arcturus to measure the radial velocity.
This was necessary, as noted above,
since field contamination toward Palomar~6
is expected to be high due to its low galactic latitude.
Once we confirmed probable cluster membership using the velocities
for each target (see Section 3.1),
we obtained useful spectra for stars A, D, and G.
We also obtained spectra of rapidly rotating early type stars in order to
remove telluric absorption features.
The journal of observations is summarized in Table~2.

The raw data reductions were performed following the similar
procedures described by Carr, Sellgren, \& Balachandran (2000).
In nodding mode, the data are stored with ``a" or ``b" name extensions
to indicate beam polarity.
Subtraction of these a-beam and b-beam images removed dark current,
residual images on the array, and any thermal or night sky line emission.
The beam images were combined together separately.
The residuals of the cosmic ray hit events on these combined images
were removed by eye using the IRAF task IMEDIT,
and then the spectrum was extracted using the IRAF task APALL.
Separate dispersion solutions were determined for the two beams
before combining them into one spectrum.
The same procedures were done for hot star spectra and then
the hot star spectra were divided into the target spectra to correct for
telluric absorption lines and for fringing in the spectra.
The target spectra were then continuum fitted and normalized.

\section{ANALYSIS}
\subsection{Radial velocity measurements}
In Figure~2, we show spectra of Arcturus and our targets
in Palomar~6 for the $^{12}$CO 2-0 bandhead centered
at $\approx$ $\lambda$ 22945 \AA.
In order to derive the heliocentric radial velocities, we cross-correlated
our object spectra with that of Arcturus
($v_{\rm r}$ = $-$5.2 km sec$^{-1}$, Evans 1967).
Table~3 shows our heliocentric radial velocity
measurements for our targets. We also list those of Minniti (1995b).
In general, our radial velocity measurements do not agree
with those of Minniti (1995b).
We believe that Minniti (1995b) misidentified star F as
a cluster member since the radial velocity of this star
is not consistent with cluster membership.

An independent check of the radial velocity for Palomar~6 remains desirable.
However, in this paper we consider our velocity measurements are correct
and stars A, C, D, and G are cluster members.
Our mean heliocentric velocity of Palomar~6
is +180.6 $\pm$ 3.2 km sec$^{-1}$ and is  20 km sec$^{-1}$ smaller
than that of Minniti (1995b).
(The error is that of the mean and is affected by both our
measurement uncertainties and by the cluster's internal
velocity dispersion.)

\subsection{Oscillator strengths}
Laboratory measurements of the oscillator strengths for
the lines in our spectral regions do not exist.
Therefore, we derived the oscillator strengths
following the procedure described by Carr et al.\ (2000),
which exploits the known abundances and stellar atmospheric
parameters for both the Sun and for Arcturus.
We used Kurucz's solar atmosphere model with $V_{\rm turb}$ = 1.0 km~s$^{-1}$
and adopted $T_{\rm eff}$ = 4300 K, $\log g$ = 1.5, and
$V_{\rm turb}$ = 1.7 km~s$^{-1}$ for Arcturus (Peterson et al.\ 1993).

We first measured the equivalent widths of the absorption lines
from the solar intensity spectrum (Wallace et al.\ 1996)
and then derived the corresponding oscillator strengths for each line
using the solar model atmosphere (Kurucz 1994) and MOOG (Sneden 1973).
For collisional damping, we considered only the van der Waals force,
and we adopted the Uns\"old approximation.
It is suspected that the collisional damping constant calculated using
the Uns\"old approximation is probably too small (e.g. Steffen 1985),
so we artificially increased the collisional damping constants by matching
the synthetic spectrum to the solar intensity spectrum of
Wallace et al.\ (1996).
To calculate the synthetic solar spectrum, we use the macroturbulent
velocity of 3.0 km~s$^{-1}$ for the Sun (Gray 1977) and the solar
elemental abundances of Anders \& Greevesse (1989).
The synthetic spectra with no enhancement in the collisional
damping constant fail to match the line core depth and
the line wing profile in particular for iron lines.
Thus we enhanced the collisional damping constant
by a factor of 3.5 for the iron lines and
by a factor of 2 for the silicon and titanium lines.
We show the synthetic solar spectra in Figure~3 and
our solar oscillator strengths in Table~4.
Note that our Fe~I oscillator strengths
are slightly different from those of Carr et al.\ (2000),
who adopted different collisional damping enhancement factors.
Their damping constants were enhanced by a factor of 3 for
Fe~I $\lambda$ 22386.90~\AA\ and
by a factor of 2 for Fe~I $\lambda$ 22399.16 \AA.
Our solar oscillator strength for Ti~I $\lambda$ 21788.63 \AA\
is rather uncertain, since it is on the wing profile of the nearby
Si~I $\lambda$ 21785.60 \AA, which has a strong wing profile
in the solar spectrum.
In the Figure, it is evident that the solar intensity spectrum
shows subtle line asymmetries, in particular for iron lines,
probably caused by the doppler shift due to the convective and
oscillatory motions in the solar atmosphere, which cannot be reproduced by
the plane-parallel one dimensional model atmosphere (Asplund 2003).

A comparison with an Arcturus spectrum provides a crucial opportunity
to examine the reliability of our solar oscillator strengths,
since the chemical composition and the stellar parameters of Arcturus are
closer to those of our target RGB stars than those of the Sun and
a high resolution, high S/N ratio spectrum is also available
for Arcturus (Hinkle et al.\ 1995).
Using the solar oscillator strengths and our equivalent width measurements
for Arcturus, we obtained [Fe/H] = $-$0.48 $\pm$ 0.07,
[Si/Fe] = $+$0.24 $\pm$ 0.09, and [Ti/Fe] = +0.61 for Arcturus.
(The errors are those of the mean.)
Our iron and silicon abundances are in good
agreement with those of Sneden et al.\ (1994),
who obtained [Fe/H] = $-$0.47 $\pm$ 0.04 and
[Si/Fe] = +0.14 $\pm$ 0.07,\footnote{
The silicon abundance for Arcturus is rather controversial.
Peterson et al.\ (1993) obtained [Fe/H] = $-$0.5 $\pm$ 0.1, [Si/Fe] = +0.4,
[Ti/Fe] = +0.3, $T_{\rm eff}$ = 4300 $\pm$ 30 K, $\log g$ = 1.5 $\pm$ 0.15, and
$V_{\rm turb}$ = 1.7 $\pm$ 0.3 km sec$^{-1}$ for Arcturus. See Table~5.}
while our titanium abundance is very high compared
to that of Sneden et al.\ (1994), [Ti/Fe] = +0.27 $\pm$ 0.04.
Figure~4 shows a comparison between the observed Arcturus spectrum
(Hinkle et al.\ 1995) and the synthetic spectrum using our solar oscillator
strengths and the elemental abundance ratios of Sneden et al.\ (1994).
The Figure clearly shows that the solar oscillator strength for
Ti~I 21788.63 \AA\ does not work for Arcturus.
The equivalent width for the Ti~I 21788.63 \AA\ line in the synthetic spectrum
is 296 m\AA\ and  it is about 66 m\AA\ smaller than that we measured
from the Arcturus spectrum of Hinkle et al.\ (1995).
We examined the atomic spectral lines of Kurucz \& Bell (1995)
and found no strong lines near Ti~I 21788.63 \AA.
We then generated the synthetic spectrum using the wavelengths and
the oscillator strengths given by Kurucz \& Bell (1995) and
measured the equivalent width produced by atomic spectral lines
around Ti~I 21788.63 \AA. We found that Ti~I 21788.25 \AA\ and
Mn~I 21788.49 \AA\ lines contribute no more than
5 m\AA\ to our equivalent measurement of Ti~I 21788.63 \AA.
We also generated a synthetic solar spectrum using the same
line list except for Ti~I 21788.25 \AA\ and
Mn~I 21788.49 \AA\ lines near Ti~I 21788.63 \AA\ and
measured the equivalent width of 0.8 m\AA\
which is too small to cause such a large
discrepancy in the titanium abundance of Arcturus.
In the synthetic solar spectrum where the spectral resolution
is much higher than that of Arcturus spectrum, the line cores
of Ti~I 21788.63 \AA\ and the combination of Ti~I 21788.25 \AA\ and
Mn~I 21788.49 \AA\ are well separated. Therefore it is not likely that
our equivalent width measurement of Ti~I 21788.63 \AA\  from the observed
solar spectrum is contaminated by the previously known
nearby atomic spectral lines.
We also examined CN molecule absorption lines (Wallace \& Hinkle 1996)
and found no known CN spectral lines at $\lambda$ 21788.63 \AA.
This leads us to suspect that there may exist unidentified atomic
spectral lines very close to Ti~I 21788.63 \AA.
Also importantly, the excitation potential energy of Ti~I 21788.63 \AA\
is 1.749 eV and this line forms higher in the atmosphere where
non-local thermodynamic equilibrium (NLTE) effects may be considered.
On the other hand, Fe~I and Si~I lines of our interest
form deep in the atmosphere due to their high excitation potential
energies and  NLTE effects to these lines are less severe.

We examined the solar $gf$ value for the Ti~I 21788.63 \AA\ line
using the giant stars I-45 and I-46 in M71.
Sneden et al.\ (1994) studied the same stars in the optical employing
high resolution echelle spectroscopy and they
obtained [Fe/H] =$-$0.78 $\pm$ 0.04, [Ti/Fe] = +0.56 $\pm$ 0.06 for I-45
and [Fe/H] = $-$0.79 $\pm$ 0.06,[Ti/Fe] = +0.47 $\pm$ 0.07 for I-46.
Using the solar oscillator strength values and the stellar parameters
of Sneden et al.\ (1994),
we obtained [Fe/H] =$-$0.9 $\pm$ 0.1, [Ti/Fe] = +1.0 for I-45 and
[Fe/H] =$-$0.8 $\pm$ 0.1, [Ti/Fe] = +0.8 for I-46.
The discrepancy in the titanium abundance is large again.
On the other hand, our silicon abundance measurement for the star I-45
is in good agreement with that of Sneden et al.\ (1994).
Our silicon abundance of the star I-46 is [Si/Fe] = $-$0.1 and
is about 0.4 dex lower than that of the star I-45.
Sneden (private communication) kindly noted that the silicon
spectral lines for the star I-46 are too weak to measure reliably
in his spectra.

If the discrepancy in the titanium abundance is caused by
NLTE effects, then using the oscillator strengths
derived from the observed Arcturus spectrum of Hinkle et al.\ (1995)
would mitigate the problem since Arcturus is a metal-rich RGB star and
the atmospheric structure of Arcturus is more similar to
our program stars than that of the Sun is.
As mentioned above, Ti~I 21788.63 \AA\ line has a low excitational
potential energy and it suffers from more severe NLTE effects
than Fe~I and Si~I lines, which have high excitational
potential energies (Ruland et al.\ 1980; Brown et al.\ 1983;
Hauschildt et al.\ 1997).

We examine the NLTE effects on our titanium abundance
using one dwarf (61 UMa) and three giants
($\alpha$ Sge, $\zeta$ Cep, and $\alpha$ Ori) with
solar metllicity.
To avoid the possible contamination by unidentified atomic spectral lines
very close to Ti~I 21788.63 \AA,
we chose three more Ti~I lines free from CN blending
at $\lambda$ 21903.31, 22216.74, 22239.03 \AA\
and we derived oscillator strengths
using the solar spectrum and the Arcturus spectrum.
For the Arcturus Ti~I oscillator strengths,
we adopt $\log n$(Ti) = 4.79 (Sneden et al.\ 1994) and
our Ti~I oscillator strengths are shown in Table~6.
Wallace \& Hinkle (1996) provided high-resolution IR spectra
($R$ $\geq$ 45,000) of 61 UMa, $\alpha$ Sge, $\zeta$ Cep, and $\alpha$ Ori
and the chemical compositions and stellar parameters
of these four stars are previously known in optical.
Table~7 shows the basic stellar parameters and iron, titanium
abundances of 61 UMa, $\alpha$ Sge, $\zeta$ Cep, and $\alpha$ Ori
(Luck 1977, Cornide \& Rego 1984, Lambert et al.\ 1984, McWilliam 1990,
Carr et al.\ 2000).
We then measured the equivalent widths for Fe~I 22386.90, 22399.16 \AA\
and Ti~I 21788.63, 21903.31, 22216.74, 22239.03 \AA\
and obtained $\log n$(Fe) and $\log n$(Ti) for the stars.
Remind that Fe~I 22386.90, 22399.16 \AA\ and Ti~I 21788.63 \AA\ lines
are of interest in our study.
Our iron abundance measurements using the solar oscillator strengths
are in excellent agreement with previous measurements in optical
within $\Delta \log n$(Fe) $\leq$ 0.05 dex.
For the titanium abundance measurements, the situation is
somewhat different.
The titanium abundance for 61 UMa is not known, but we
assume that this thin disk, solar-type star has [Ti/Fe] = 0.00,
from which we deduce log $n$(Ti) = 4.99. As Table~7 shows,
using $gf$ values for the different Ti~I lines derived
from solar spectrum leads to consistent titanium abundances
for 61 UMa, whose surface gravity is similar to the Sun.
But the solar Ti~I $gf$ values do not appear to yield
correct abundances for the three giant stars.
Table~7 shows that in those cases, the $gf$ values
derived from the Arcturus spectrum provide a better match.
This may reflect unknown problems in either underlying
and unidentified absorption lines or, since the behavior
appears to affect all the Ti~I lines, problems in the
adopted $T-\tau$ relations for giants compared to dwarfs.

For our titanium abundance determination in Palomar~6, in which we
rely on giant stars, we therefore adopt $gf$ values
derived from the Arcturus spectrum. Table~8 summarizes our
adopted $gf$ values.

\subsection{Stellar parameters and model atmospheres}
\subsubsection{The effective temperature}
Having good stellar parameters, such as the effective temperature and
the surface gravity, is very important in any stellar abundance study,
since the absolute or the relative elemental abundance scales
will depend on the input stellar parameters.
In optical spectroscopy, the effective temperature is usually
determined by using weak neutral iron lines with
log(W$_{\lambda}$/$\lambda$) $\lesssim$ $-5.2$,
lying on the linear part of curve of growth (COG).
Since only the two very strong neutral iron lines with similar excitation
potentials are available in our spectral regions,
we have to rely upon photometric temperature estimates.
However, applying the photometric method using relations between
color versus $T_{\rm eff}$ for Palomar~6 still poses a problem,
since the interstellar reddening value for the cluster is not well known.
In the case of using the $(B-V)$ color index to estimate
the stellar temperature, an uncertainty of 0.1 mag in $E(B-V)$ will result in
an uncertainty of $\approx$ 100 K in the photometric surface temperature
in the stellar temperature range of our interest.
The discrepancy between previous interstellar reddening estimates
for Palomar~6 is as large as $\Delta E(B-V)$ $\approx$ 0.15 mag
(Lee \& Carney 2002a).
We, therefore, devise an alternative photometric temperature determination
method, which is independent of interstellar reddening
and the photometric zero points, using the $K$ magnitude differences
between the horizontal branch (HB) stars and the RGB stars.

In Figure~5, we show a plot of $T_{\rm eff}$ versus $\log L/L_\odot$
for the RGB sequences with Z = 0.010, 0.004
(equivalently [Fe/H] = $-$0.5, $-$0.9)
and the zero-age horizontal branch (ZAHB; Kim et al. 2002;
Yi, Demarque, \& Kim 1997).
The differences in $\log L/L_\odot$
(i.e., the differences in bolometric magnitudes)
between HB and RGB stars are linearly related to
the surface temperature of the RGB stars,
with a slight dependence on metallicity.
Since the bolometric correction for RGB stars in the $K$ passband, $BC_K$,
is linearly correlated with the effective temperature
(see, for example, Bessell, Castelli, \& Plez 1998),
the difference in the $K$ magnitude is also linearly correlated with
the difference in luminosity, $\log L/L_\odot$.
Therefore, we can estimate the temperature of our target RGB stars
by simply measuring the difference in $K$ magnitude between the HB and
the target RGB stars, $\Delta (K_{HB}-K)$.
Since HB stars are not located horizontally in IR CMDs,
it is necessary to adopt the $K$ magnitude at the intersection
between RGB and red HB (RHB) in IR photometry for our HB reference level,
$K_{(RGB,RHB)}$ (for example, see Kuchinski et al.\ 1995).
In practice, instead of using the theoretical values,
we relied upon the $T_{\rm eff}$ versus $\Delta (K_{(RGB,RHB)}-K)$ diagram
for RGB stars in well-studied globular clusters with similar metallicities
as Palomar~6.

We show $JK$ CMDs of the metal-rich globular clusters
47~Tuc ([Fe/H] = $-$0.7, Frogel, Persson, \& Cohen 1981),
M69 ([Fe/H] = $-$0.7, Frogel, Persson, \& Cohen 1983; Davidge \& Simons 1991),
M71 ([Fe/H] = $-$0.7, Frogel, Persson, \& Cohen 1979),and
M107 ([Fe/H] = $-$1.0, Frogel, Persson, \& Cohen 1983) in Figure~6.
We adopt $K_{(RGB,RHB)}$ magnitudes from Kuchinski et al.\ (1995)
for these template clusters and we use  $K_{(RGB,RHB)}$  = 13.53 mag for
Palomar~6 (see Lee \& Carney 2002a).
Since the interstellar reddening values are well known for template
clusters, we can derive the effective temperature for the RGB stars
in template clusters using the relation given by
Alonso, Arribas, \& Martinez-Roger (1999).
In Figure~7, we show a plot of $\Delta (K_{(RGB,RHB)}-K)$ versus
$T_{\rm eff}$ for the  RGB stars in the template clusters and we obtain
a linear relation between $T_{\rm eff}$ and $\Delta (K_{(RGB,RHB)}-K)$,
\begin{equation}
T_{\rm eff} = -192.1(\pm 9.5)\times\Delta (K_{(RGB,RHB)}-K) + 4639.2(\pm 28.5).
\end{equation}
Using this relation and $\Delta (K_{(RGB,RHB)}-K)$ values for our target RGB
stars in Palomar~6 listed in Table~1, we obtain
$T_{\rm eff}$ = 3718, 3741, and 3667 K for stars A, D, and G, respectively.
The effective temperature estimates for Palomar~6 A, D, and G
using the $(V-K)$ versus $T_{\rm eff}$ relation
given by Alonso et al.\ (1999)
are 3703, 3747, and 3597 K for $E(B-V)$ = 1.3 mag (Lee \& Carney 2002a).
These values agree with those from the
$\Delta (K_{(RGB,RHB)}-K)$ versus $T_{\rm eff}$ relation
within $\Delta T_{\rm eff}$ $\approx$ 100 K.
As a cross-check, we also obtained the surface temperature for M71 RGB stars
using equation~(1) and we have $T_{\rm eff}$ =4043 K for I-45
and $T_{\rm eff}$ = 4019 K for I-46.
Our surface temperatures for M71 RGB stars are in good agreement
with those of Sneden et al.\ (1994), who obtained temperatures
of 4050 K and 4000 K, respectively.

\subsubsection{The surface gravity}
To estimate the surface gravity, we rely on the model isochrones of
Yi, Demarque, \& Kim (1997) and Kim et al. (2002).
First we calculated the bolometric correction in the $K$ passband, $BC_K$,
for the temperatures and surface gravities of each grid of
model isochrones by interpolating the ($T_{\rm eff}$, $\log g$, $BC_K$)
grids of Bessell, Castelli, \& Plez (1998).
Then we calculated the bolometric magnitudes ($M_{bol}$) using $\log L$ and
obtained the magnitudes in $K$ passband ($M_K$)
by subtracting $BC_K$ from $M_{bol}$
for each grid of the model isochrones, assuming
\begin{equation}
M_K = M_{bol} - BC_K.
\end{equation}
This procedure allows us to calculate $\Delta (K_{(RGB,RHB)}-K)$ for
the model isochrones.
In Figure~8, we show a plot of $\Delta (K_{(RGB,RHB)}-K)$ versus $\log g$
for the model isochrones with [Fe/H] = $-$0.9, $-$0.5 and
[$\alpha$/Fe] = +0.3 for 10, 14 Gyr.
In the Figure, we also show the locations of RGB stars in M71 and
our program stars in Palomar~6.
For M71 RGB stars, we obtained $\log g$  = 1.0 for I-45 and I-46.
Our surface gravities for M71 RGB stars are $\Delta\log g$  = 0.2 dex larger
than those of Sneden et al.\ (1994).
For Palomar~6, we obtained $\log g$ = 0.3 for Palomar~6 A and D, and
$\log g$ = 0.2 for Palomar~6 G assuming [Fe/H] $\approx$ $-$1.0 for
Palomar~6 (see Lee \& Carney 2002a).
Note that our surface gravity estimates using the model isochrones
should be considered as the upper limit of surface gravities
of our program stars, since the mass loss has not been included
in model isochrone calculations (Kim, Y. -C., private communication).
Since our program stars in Palomar~6 lie near the tip of RGB,
the mass loss rate will be large, leading to lower surface gravities.

\subsubsection{The microtubulent velocity}
To estimate the microturbulent velocity $V_{\rm turb}$ requires
many spectral lines, preferably Fe~I lines in optical spectroscopy,
both on the linear and the flat parts of the COG.
Since we have only two strong Fe~I (or two Si~I) lines and we are not able
to properly estimate the microturbulent velocities for our program RGB stars.
Therefore we adopt the microturbulent velocity of 2 km sec$^{-1}$
for our program RGB stars without any verification.
In Figure~9, we show comparisons of
$\log g$ versus $T_{\rm eff}$ versus $V_{\rm turb}$
of RGB stars in globular clusters extensively studied in optical
(Brown, Wallerstein, \& Oke 1990; Ivans et al. 1999, 2001;
Norris \& Da Costa 1995; Sneden et al. 1994, 2004).
Nine stars in these six clusters have $\log g$ $\leq$ 0.3,
comparable to our program stars in Palomar~6.
Those nine stars have a mean $V_{\rm turb}$ of 1.9 $\pm$ 0.2 km sec$^{-1}$.
If we include stars with $\log g$ $\leq$ 0.5, there are 23 stars
and they have a mean $V_{\rm turb}$ of 2.0 $\pm$ 0.3 km sec$^{-1}$
(The error is the standard deviation).
Since these stars are very similar to our program stars in Palomar~6,
our choice of $V_{\rm turb}$ = 2.0 km sec$^{-1}$ is thought to
be reasonable.
It should be noted that, however, the microturbulent velocity appears
to increase with decreasing surface gravity in M supergiants
(see, for example, Figure~4 of Ram\'\i rez et al.\ 2000).

\subsubsection{The model atmosphere}
With temperature and surface gravity estimates,
72-depth plane-parallel local thermodynamic equilibrium (LTE)
model atmospheres were computed using the program ATLAS9,
written and supplied by Dr.\ R.\ L.\ Kurucz.
The model atmospheres were computed using opacity distribution functions
and abundances with enhanced abundances of all the ``$\alpha$" elements
(O, Ne, Mg, Si, S, Ar, Ca, and Ti) by 0.4~dex,
assuming our target RGB stars would prove to be
$\alpha$-element enhanced (see discussions below).
The abundance analysis was performed using the program MOOG (Sneden 1973).

Since our abundance analysis is based upon the neutral lines,
the elemental abundances are not sensitive to the surface gravity.
On the other hand, since we rely on very strong lines,
our results will depend on the choice of
microturbulent velocity of the model atmosphere,
which we are not able to estimate using our target spectra.

\section{RESULTS}
We show our equivalent width measurements in Table~9 and
the elemental abundances for Palomar~6 and M71 in Table~10.
We also show the elemental abundances of Sneden et al.\ (1994)
in the Table.
In Table~11, we show estimated errors resulting from
the uncertainties in  the input model atmosphere $\delta T_{\rm eff}$ = +200 K,
$\delta \log g$ = $-$0.2, and $\delta V_{\rm turb}$ = +0.5 km sec$^{-1}$,
which are appropriate for our analysis.
In general, [Fe/H] and [Si/Fe] ratios are not sensitive
to the changes in the input parameters of model atmospheres.
However, [Ti/Fe] is very sensitive to the surface temperature and
the microturbulent velocity.
An uncertainty of $\delta T_{\rm eff}$ = +200 K leads to an uncertainty of
$\approx$ $+$0.3 dex and an uncertainty of
$\delta V_{\rm turb}$ = +0.5 km sec$^{-1}$
leads to $\approx$ $-$0.3 dex in the [Ti/Fe] ratio.

Table~10 shows that our M71 abundances are consistent with those obtained
by Sneden et al., given the uncertainties and the limited number
of lines available for our analyses.

The mean metallicity for Palomar~6 is [Fe/H] = $-$1.0 $\pm$ 0.1,
consistent with that derived from the slope of the RGB in the $JK$ CMD,
[Fe/H] $\approx$ $-$1.1 (Lee \& Carney 2002a),
and is slightly more metal-poor than M71,
[Fe/H] = $-$0.8 $\pm$ 0.1, unless our surface temperatures or
the microturbulent velocities for the Palomar~6 RGB stars are
inaccurate by more than $\Delta T_{\rm eff} \approx$ 400 K or
$\Delta V_{\rm turb} \approx$ $-$0.7 km sec$^{-1}$.
We show the synthetic spectra for [Fe/H] = $-$1.0, $-$0.5, 0.0,
and the observed spectra of the Palomar~6 RGB stars in Figure~10.
In the Figure we used an ad hoc $gf$ value for
Sc~I $\lambda$ 22400.70 \AA\ since the hyperfine splitting components
for the line are not known.
The Figure clearly indicates that the observed Fe~I absorption lines
are too weak  to lead to the super- or near-solar metallicity as claimed by
Minniti (1995b) or Bica et al.\ (1998).
Figure~11 shows synthetic spectra for the star Palomar~6 G
using the model atmospheres with [Fe/H] = +0.20,
which is the metallicity of Palomar~6 claimed by Minniti (1995).
As can be seen in the Figure, synthetic spectra do not reproduce
the observed spectrum of the star Palomar~6 G until
$T_{\rm eff}$ $\gtrsim$ 5000 K,
which is an unreasonably high surface temperature
for the super-metal-rich RGB star (see also Figure~6).
Therefore, it is most likely that Palomar~6 is much more metal-poor than
claimed by  Minniti (1995b) or Bica et al.\ (1998).

Our [Si/Fe] ratio for the star I-45 shows good agreement
with the results of Sneden et al.\ (1994), suggesting that our measurement
for the silicon abundance is reliable.
It appears that the silicon abundance of the star Palomar~6 G
is enhanced by $\approx$ 0.4 dex.
The titanium abundance is enhanced by
$\approx$ 0.5 dex relative to the iron abundance.
Figure~12 shows synthetic spectra of the star Palomar~6 G
with [Si/Fe] = +0.4 and [Ti/Fe] = +0.5.
In their Figure~9, Lee \& Carney (2002a) showed the Palomar~6 $VI$ CMD
drooping at the RGB-tip, which led Ortolani et al.\ (1995) to conclude that
Palomar~6 has [Fe/H] $\approx$ $-$0.4.
The droop seen at the tip of the RGB branch in the CMD of Palomar~6
could be due to its high titanium abundance
which increases the line blanketing
effect by TiO bands and may not due to the overall high metallicity.
In particular, since the star Palomar~6 G has a low effective temperature,
this effect would be large.
It should be emphasized again that our [Ti/Fe] ratio measurement
is very sensitive to $T_{\rm eff}$ and $V_{\rm turb}$ of
the input model atmosphere and our derived titanium abundance
for Palomar~6 relies on the single line measurement in a single star.
As shown in Table~11, an uncertainty of $\Delta T_{\rm eff}$ = 200~K leads to
the uncertainty of almost 0.3 dex in the [Ti/Fe] ratio.
Our titanium abundance of Palomar~6 relied upon
a very strong single line measurement and the microturbulent velocity
was not measured in an appropriate way.
A change of 0.5 km sec$^{-1}$ in the microturbulent velocity results in
$\approx$ $-$0.3 dex change in the [Ti/Fe] ratio.

\section{DISCUSSIONS}
\subsection{Is Palomar~6 part of the inner halo or the bulge populations?}
Globular clusters do not define a single homogeneous population with
a single history and, in particular, the Galactic central regions may
contain globular clusters belonging to various populations.
The globular clusters near the Galactic center may provide the means
by which we can improve  our understanding of the processes and chronology
involved in the formation of the Galactic halo and the bulge.
Carney, Latham, \& Laird (1990) and Wyse \& Gilmore (1992) argued that
the halo is the ancestor to the bulge.
The idea is that the halo is metal-poor because star formation terminated
early by the loss of gas from the halo, and the low angular momentum
of the halo means the gas should have largely collected dissipatively
into the central regions (i.e., into the bulge).
However, the definition of the bulge clusters currently being used is
rather ambiguous mainly due to lack of complete kinematical information.
Minniti (1995a) drew attention to a set of clusters that may
be associated with the Galaxy's bulge, distinct from the central
concentration of the spheroidal halo clusters.
In practice, the metal-rich clusters located within a few kpc
from the Galactic center are usually categorized as the bulge population,
but halo clusters formed in the metal-rich tail may also exist
in the Galactic central regions.
In a recent study, Dinescu et al.\ (2003) discussed
the population status of seven globular clusters
with $R_{GC}$ $\lesssim$ 3.2 kpc based on their actual space velocities.
One of their intriguing findings is that the metal-rich globular cluster
NGC~6316 is most likely a halo cluster, in spite of its high
metallicity, [Fe/H] = $-$0.55.
Unfortunately, Palomar~6 does not as yet have a
measurement of its proper motion. To assign it to the disk, bulge,
or halo populations, we will have to rely only
upon its metallicity, position, and radial velocity.

In Figure~13, we show the metallicity distribution and
the kinematical properties of globular clusters within 3 kpc from
the Galactic center using data available from Harris (1996).
In addition, we adopt the recent metallicity measurements by
Origlia, Rich, \& Castro (2002) for Liller~1,
Carretta et al.\ (2001) for NGC~6528, and Cohen et al.\ (1999) for NGC~6553.
The metallicity distribution shows two peaks at [Fe/H] $\approx$ $-$1.3
and $-$0.4, suggesting that at least two distinct populations exist
in the Galactic central regions.
The bottom panel shows a plot of $V_S$ versus $\cos \psi$ for
the metal-rich and metal-poor clusters using the relations given by
Frenk \& White (1980), corrected by Zinn (1985),
where $V_S$ is the radial velocity observed at the Sun's position by an
observer at rest with respect to the galactic center and
$\psi$ is the angle between the line of sight and the direction of
Galactic rotation at the cluster (Zinn 1993).
The $\cos \psi$ and $V_S$ values are listed in Table~13.
In the Figure, the mean rotation velocity, $\langle V_{\rm rot}\rangle$,
is given by the slope of the straight line,
and the line-of-sight velocity dispersion, $\sigma_{los}$, is given by
the standard deviation of the points about the line.
The fit for the metal-poor clusters ([Fe/H] $<$ $-$1.0) is represented
by the solid line and that for the metal-rich clusters ([Fe/H] $>$ $-$1.0)
by the dashed line. The metal-poor clusters have
$\langle V_{\rm rot}\rangle$ = +111 $\pm$ 57 km~sec$^{-1}$ and
$\sigma_{los}$ = 137 $\pm$ 31 km~sec$^{-1}$,
while the metal-rich clusters have
$\langle V_{\rm rot}\rangle$ = +89 $\pm$ 36 km~sec$^{-1}$ and
$\sigma_{los}$ = 99 $\pm$ 23 km~sec$^{-1}$.
The rotational velocity solution of the metal-rich clusters with
the lower mass bar-like kinematics is also shown in the Figure
by the dotted line (Burkert \& Smith 1997).
There is no distinctive difference in mean kinematics
between the two metallicity regimes, but the metallicity
distribution shows that if there are two different groups
based on different mean metallicities, the overlap in
[Fe/H] is considerable. We can, nonetheless, still employ
kinematics as a discriminant since more disk-like or
bulge-like clusters will, on average be closer to the
mean rotational solutions than would halo clusters.
The observed radial velocity of Palomar~6 deviates by over
150 km~sec$^{-1}$ from the rotational velocity solutions,
suggesting that Palomar~6 may have halo kinematics.
We believe that Palomar~6 is probably a globular cluster
that is more probably classified as belonging to the halo population
rather than a bulge or disk population, based on both its kinematics
and its relatively low metallicity.
Like other halo clusters, its high [$\alpha$/Fe] ratio suggests
that the star formation rate was very high when the cluster formed.
Additional spectroscopic studies of high and low metallicity
and high and low velocity clusters in the inner Galaxy will be
needed to unravel the complex histories of star formation.

\subsection{The silicon and the titanium abundances of Palomar~6}
As we discussed above, the measurement of $\alpha$-element abundances
of globular clusters provides important information regarding relative ages,
in the sense that the enhanced value of [$\alpha$/Fe] ratios indicates
the high star formation rate and old age of the system.
In Figure~14, we show the silicon and the titanium abundances
of globular clusters as a function of metallicity
(see Lee \& Carney 2002b and references therein).
In the Figure, our [Ti/Fe] ratio of Palomar~6 appears to be
0.2 dex higher than the mean value of [Ti/Fe] of other globular
clusters, [Ti/Fe] $\approx$ 0.3 dex.
However, given the uncertainty of Palomar~6 [Ti/Fe] value,
the significance of this difference is hard to access without
further data.
The enhanced $\alpha$-element abundances of Palomar~6 suggest that
Palomar~6 must have formed very quickly,
in spite of its high metallicity.
If true, the Galactic inner halo must have experienced a very rapid
chemical enrichment history.

Figure~15 shows [Si/Ti] ratio versus Galactocentric distance,
$R_{GC}$, of globular clusters.
The bisector linear fit to the old halo clusters (18 clusters) and
that to the old halo clusters with $R_{GC}$ $\leq$ 8 kpc (12 clusters)
are also shown in the Figure.
As we mentioned above, Lee \& Carney (2002b) claimed that
[Si/Ti] ratios of the old halo globular clusters are anticorrelated with
Galactocentric distances, in the sense that the silicon abundance
increases and the titanium abundance decreases with Galactocentric distances.
They argued that the contributions from SNe~II events with different
progenitor masses could explain such a gradient.
In the Figure, the [Si/Ti] ratio of Palomar~6 does not appear
to follow the trend claimed by Lee \& Carney (2002b) and, further,
the [Si/Ti] ratio of Palomar~6 appears to be lower than those of metal-rich
bulge clusters NGC~6528 ([Fe/H] = +0.07, Carretta et al.\ 2001) and
NGC~6553 ([Fe/H] = $-$0.16, Cohen et al.\ 1999).
The [Si/Ti] ratio of Palomar~6 appears to agree with
those of the metal-rich bulge K giants by McWilliam \& Rich (1994).
This may suggest that the chemical enrichment history of Palomar~6
was more complex than those of other clusters
in the Galactic central regions.
Again, given the uncertainty in the Palomar~6 [Si/Ti] value,
the significance of this difference is hard to assess without
further data.

\section{SUMMARY}
In this paper, we have discussed high-resolution infrared echelle
spectroscopy for the globular clusters Palomar~6 and M71.
Our mean heliocentric velocity of Palomar~6
is +180.6 $\pm$ 3.2 km sec$^{-1}$ and is  20 km sec$^{-1}$ smaller
than that of Minniti (1995b).
Contrary to the recent results using low resolution spectroscopy,
our results have shown that Palomar~6 has an intermediate
metallicity with  [Fe/H] = $-$1.0 $\pm$ 0.1
and is slightly more metal-poor than M71.
Our estimated uncertainties in the surface temperatures,
gravities, and microturbulent velocities for the
three stars we have studied will not affect our derived
[Fe/H] value by more than 0.2 dex. We believe that
previous estimates, based on lower resolution, lower-S/N
spectra were affected by unrecognized molecular line
blanketing or by stars that are not members of the cluster.
The [$\alpha$/Fe] ratio of Palomar~6 appears to be enhanced
by $\approx$ 0.3 -- 0.5 dex in spite of its high metallicity,
suggesting that the Galactic inner halo must have experienced a very rapid
chemical enrichment history.

\acknowledgments
JWL wishes to thank J. S. Carr, C. Sneden and Y. -C. Kim
for their kind discussions.
We thank an anonymous referee for useful comments and
a careful review of the paper.
This research was supported by the National Aeronautics and Space
Administration (NASA) grant number GO-07318.04-96A from the Space Telescope
Science Institute, which is operated by the Association of Universities
for Research in Astronomy (AURA), Inc., under NASA contract NAS 5-26555.
We also thank the National Science Foundation for financial support via
grants AST$-$9619381, AST$-$9888156 and AST$-$030541
to the University of North Carolina.
Support for this work was also provided by the Korea Science
and Engineering Foundation (KOSEF) to the Astrophysical Research Center
for the Structure and Evolution of the Cosmos (ARCSEC).
SBC is pleased to acknowledge NSF grants 98-19870 and 00-98619
to the University of Maryland.
\clearpage

\clearpage

\begin{deluxetable}{crcccc}
\tablecaption{Palomar~6 program stars.}
\tablenum{1}
\tablewidth{0pc}
\tablehead{
\colhead{Id.\tablenotemark{1}} &
\colhead{$K$} &
\colhead{$J-K$} &
\colhead{$H-K$} &

\colhead{$V$\tablenotemark{2}} &
\colhead{$\Delta(K_{(RGB,RHB)}-K)$} }
\startdata
A &  8.733 &  1.543 & 0.262 & 16.94 & 4.797 \\
B &  8.541 &  1.482 & 0.269 & 16.61 & 4.989 \\
C &  9.770 &  1.456 & 0.262 & 17.11 & 3.760 \\
D &  8.855 &  1.551 & 0.284 & 16.89 & 4.675 \\
E &  9.632 &  1.462 & 0.249 & 17.18 & 3.898 \\
F &  8.267 &  1.638 & 0.322 & 17.22 & 5.263 \\
G &  8.467 &  1.646 & 0.315 & 17.27 & 5.063 \\
\enddata
\tablenotetext{1}{Minniti (1995).}
\tablenotetext{2}{Ortolani et al.\ (1995).}
\end{deluxetable}

\clearpage

\begin{deluxetable}{clccccc}
\tablecaption{Journal of observations.}
\tablenum{2}
\tablewidth{0pc}
\tablehead{
\multicolumn{2}{c}{Id.} &
\colhead{$\lambda$ (\AA)} &
\colhead{Date (UT)} &
\colhead{$t_{exp}$ (sec)} &
\colhead{S/N}}
\startdata
Pal 6 & A & 22395 & 24 May 1998 & 8400 & 45 \\
      & D & 22395 & 25 May 1998 & 4200 & 45 \\
      & G & 22395 & 24 May 1998 & 5400 & 60 \\
      & G & 21815 & 25 May 1998 & 4200 & 70 \\
 & & & & & \\
M71  & I-45 & 22395 & 14 July 1997 & 2100 & 40 \\
     & I-45 & 21815 & 14 July 1997 & 2400 & 50 \\
     & I-46 & 22395 & 13 July 1997 & 2700 & 55 \\
     & I-46 & 21815 & 13 July 1997 & 2400 & 50 \\
\enddata
\end{deluxetable}

\clearpage

\begin{deluxetable}{ccccc}
\tablecaption{Heliocentric radial velocities of Palomar~6 program stars.}
\tablenum{3}
\tablewidth{0pc}
\tablehead{
\colhead{Id.} &
\colhead{$v_{\rm r}$\tablenotemark{1}} &
\colhead{$\sigma$\tablenotemark{1}} &
\colhead{$v_{\rm r}$\tablenotemark{2}} &
\colhead{Membership} \\
\colhead{} &
\colhead{(km sec$^{-1}$)} &
\colhead{(km sec$^{-1}$)} &
\colhead{(km sec$^{-1}$)} &
\colhead{} }
\startdata
A & 185.3 & 4.0 & \nodata  & Yes \\
B & ~26.7 & 1.7 & 56.4 & No \\
C & 173.5 & 3.6 & 243.3 & Yes \\
D & 186.8 & 2.3 & 201.2 & Yes \\
E & $-$13.5 & 2.1 &  \nodata & No \\
F & 134.5 & 3.3 & 200.0 & No \\
G & 176.7 & 2.2 & \nodata & Yes \\
 & & & \\
Mean & 180.6 & & 200.6 & \\
\enddata
\tablenotetext{1}{This work.}
\tablenotetext{2}{Minniti (1995).}
\end{deluxetable}

\clearpage

\begin{deluxetable}{ccccrrcc}
\tablecaption{Solar oscillator strengths.}
\tablenum{4}
\tablewidth{0pc}
\tablehead{
\multicolumn{3}{c}{} &
\multicolumn{1}{c}{} &
\multicolumn{2}{c}{Sun} &
\multicolumn{1}{c}{} &
\multicolumn{1}{c}{Carr et al.}\\
\cline{5-6}\\
\colhead{$\lambda$} &
\colhead{Ele.} &
\colhead{E.P.} &
\colhead{} &
\colhead{EW} &
\colhead{$\log gf$} &
\colhead{} &
\colhead{} \\
\colhead{(\AA)} &
\colhead{} &
\colhead{(eV)} &
\colhead{} &
\colhead{(m\AA)} &
\colhead{} &
\colhead{} &
\colhead{} }
\startdata
22386.90 & Fe~I & 5.033 && 229 & $-$0.490 &&  $-$0.481\\
22399.16 & Fe~I & 5.099 && ~72 & $-$1.275 &&  $-$1.249\\
21785.60 & Si~I & 6.719 && 400 &  ~~0.295 &&  \nodata \\
21825.63 & Si~I & 6.721 && 293 &  ~~0.030 &&  \nodata \\
21788.63 & Ti~I & 1.749 && ~33 & $-$1.265 &&  \nodata \\
\enddata
\end{deluxetable}

\clearpage

\begin{deluxetable}{clcc}
\tablecaption{Comparisons of elemental abundances of Arcturus.}
\tablenum{5}
\tablewidth{0pc}
\tablehead{
\colhead{} &
\colhead{This work} &
\colhead{Sneden et al.} &
\colhead{Peterson et al.} }
\startdata
[Fe/H]  & $-$0.48 $\pm$ 0.07  & $-$0.47  &  $-$0.5 \\

[Si/Fe] & $+$0.24 $\pm$ 0.09   &  $+$0.14   &   $+$0.4  \\

[Ti/Fe] & $+$0.61              &  $+$0.27   &   $+$0.3  \\
\enddata
\end{deluxetable}

\clearpage

\begin{deluxetable}{ccccrcccr}
\tablecaption{Comparisons of Ti~I oscillator strengths.}
\tablenum{6}
\tablewidth{0pc}
\tablehead{
\multicolumn{3}{c}{} &
\multicolumn{2}{c}{Sun} &
\multicolumn{1}{c}{} &
\multicolumn{3}{c}{Arcturus} \\
\cline{4-5}\cline{7-9}\\
\colhead{$\lambda$} &
\colhead{E.P.} &
\colhead{} &
\colhead{EW} &
\colhead{$\log gf$} &
\colhead{} &
\colhead{EW} &
\colhead{$\log n$(Ti~I)\tablenotemark{1}} &
\colhead{$\log gf$} \\
\colhead{(\AA)} &
\colhead{(eV)} &
\colhead{} &
\colhead{(m\AA)} &
\colhead{} &
\colhead{} &
\colhead{(m\AA)} &
\colhead{} &
\colhead{} }
\startdata
21788.63 &  1.749 && 33 & $-$1.265 && 361 & 5.17 & $-$0.884 \\
21903.31 &  1.739 && 17 & $-$1.600 && 316 & 5.18 & $-$1.210 \\
22216.74 &  1.734 && 10 & $-$1.850 && 237 & 4.91 & $-$1.730 \\
22239.03 &  1.739 && 11 & $-$1.810 && 264 & 5.05 & $-$1.555 \\
\enddata
\tablenotetext{1}{Ti~I abundances using the solar Ti~I $gf$ values
in col. 4.}
\end{deluxetable}

\clearpage

\begin{landscape}
\begin{deluxetable}{ccccccccccccccccc}
\tablecaption{Comparisons of Ti~I oscillator strengths.}
\tabletypesize{\scriptsize}
\tablenum{7}
\tablewidth{0pc}
\tablehead{
\multicolumn{2}{c}{} &
\multicolumn{3}{c}{61 UMa ~(G8 V)} &
\multicolumn{1}{c}{} &
\multicolumn{3}{c}{$\alpha$ Sge ~(G1 III)} &
\multicolumn{1}{c}{} &
\multicolumn{3}{c}{$\zeta$ Cep ~(K1.5 Ib)} &
\multicolumn{1}{c}{} &
\multicolumn{3}{c}{$\alpha$ Ori ~(M2 Ia)} \\
\cline{3-5}\cline{7-9}\cline{11-13}\cline{15-17}\\
\multicolumn{2}{c}{$T_{\rm eff}$} &
\multicolumn{3}{c}{5220\tablenotemark{1}} &
\multicolumn{1}{c}{} &
\multicolumn{3}{c}{5440\tablenotemark{2}} &
\multicolumn{1}{c}{} &
\multicolumn{3}{c}{4400\tablenotemark{3}} &
\multicolumn{1}{c}{} &
\multicolumn{3}{c}{3540\tablenotemark{4}} \\
\multicolumn{2}{c}{(K)} &
\multicolumn{15}{c}{} \\
\multicolumn{2}{c}{$\log g$} &
\multicolumn{3}{c}{4.7\tablenotemark{1}} &
\multicolumn{1}{c}{} &
\multicolumn{3}{c}{3.1\tablenotemark{2}} &
\multicolumn{1}{c}{} &
\multicolumn{3}{c}{1.0\tablenotemark{3}} &
\multicolumn{1}{c}{} &
\multicolumn{3}{c}{0.0\tablenotemark{4}} \\
\multicolumn{2}{c}{$V_{\rm turb}$} &
\multicolumn{3}{c}{1.4\tablenotemark{1}} &
\multicolumn{1}{c}{} &
\multicolumn{3}{c}{3.1\tablenotemark{2}}  &
\multicolumn{1}{c}{} &
\multicolumn{3}{c}{2.6\tablenotemark{3}} &
\multicolumn{1}{c}{} &
\multicolumn{3}{c}{3.2\tablenotemark{4}} \\
\multicolumn{2}{c}{(km sec${-1}$)} &
\multicolumn{15}{c}{} \\
\multicolumn{2}{c}{$\log n$(Fe~I)} &
\multicolumn{3}{c}{7.55\tablenotemark{1}} &
\multicolumn{1}{c}{} &
\multicolumn{3}{c}{7.37\tablenotemark{2}} &
\multicolumn{1}{c}{} &
\multicolumn{3}{c}{7.75\tablenotemark{3}} &
\multicolumn{1}{c}{} &
\multicolumn{3}{c}{7.52\tablenotemark{4}} \\
\multicolumn{2}{c}{$\log n$(Ti~I)} &
\multicolumn{3}{c}{\nodata} &
\multicolumn{1}{c}{} &
\multicolumn{3}{c}{4.87\tablenotemark{2}} &
\multicolumn{1}{c}{} &
\multicolumn{3}{c}{5.31\tablenotemark{3}} &
\multicolumn{1}{c}{} &
\multicolumn{3}{c}{5.07\tablenotemark{5}} \\
\hline\\
\colhead{$\lambda$} &
\colhead{Id} &
\colhead{EW} &
\colhead{$\log n$} &
\colhead{$\log n$} &
\colhead{} &
\colhead{EW} &
\colhead{$\log n$} &
\colhead{$\log n$} &
\colhead{} &
\colhead{EW} &
\colhead{$\log n$} &
\colhead{$\log n$} &
\colhead{} &
\colhead{EW} &
\colhead{$\log n$} &
\colhead{$\log n$} \\
\colhead{(\AA)} &
\colhead{} &
\colhead{(m\AA)} &
\colhead{Sol. $gf$} &
\colhead{Arc. $gf$} &
\colhead{} &
\colhead{(m\AA)} &
\colhead{Sol. $gf$} &
\colhead{Arc. $gf$} &
\colhead{} &
\colhead{(m\AA)} &
\colhead{Sol. $gf$} &
\colhead{Arc. $gf$} &
\colhead{} &
\colhead{(m\AA)} &
\colhead{Sol. $gf$} &
\colhead{Arc. $gf$}}

\startdata
22386.90 & Fe~I & 267     & 7.52    & \nodata &&
                  257     & 7.29    & \nodata &&
                  503     & 7.79    & \nodata &&
                  577     & 7.54    & \nodata \\
22399.16 & Fe~I & 105     & 7.56    & \nodata &&
                  126     & 7.50    & \nodata &&
                  320     & 7.76    & \nodata &&
                  352     & 7.42    & \nodata \\
21788.63 & Ti~I & ~66     & 4.97    & 4.59    &&
                  122     & 5.19    & 4.81    &&
                  585     & 5.65    & 5.27    &&
                  919     & 5.42    & 5.04    \\
21903.31 & Ti~I & ~41     & 5.02    & 4.63    &&
                  \nodata & \nodata & \nodata &&
                  518     & 5.71    & 5.32    &&
                  \nodata & \nodata & \nodata \\
22216.74 & Ti~I & \nodata & \nodata & \nodata &&
                  \nodata & \nodata & \nodata &&
                  398     & 5.50    & 5.38    &&
                  769     & 5.31    & 5.19    \\
22239.03 & Ti~I & ~32     & 5.10    & 4.84    &&
                  \nodata & \nodata & \nodata &&
                  433     & 5.60    & 5.34    &&
                  792     & 5.37    & 5.12    \\
\hline\\
& $\langle$ Fe~I $\rangle$ && 7.54 & \nodata &&
                  & 7.40 &\nodata &&
                  & 7.78 &\nodata &&
                  & 7.48 &\nodata \\
& $\langle$ Ti~I $\rangle$ && 5.03 & 4.68 &&
                  & 5.19 & 4.81 &&
                  & 5.62 & 5.33 &&
                  & 5.37 & 5.12 \\
\enddata
\tablenotetext{1}{Cornide \& Rego (1984).}
\tablenotetext{2}{McWilliam (1990).}
\tablenotetext{3}{Luck (1977).}
\tablenotetext{4}{Carr et al.\ (2000).}
\tablenotetext{5}{Lambert et al.\ (1984).}
\end{deluxetable}
\end{landscape}

\clearpage

\begin{deluxetable}{ccccl}
\tablecaption{Adopted oscillator strengths.}
\tablenum{8}
\tablewidth{0pc}
\tablehead{
\colhead{$\lambda$} &
\colhead{Ele.} &
\colhead{E.P.} &

\colhead{$\log gf$} &
\colhead{Note} \\
\colhead{(\AA)} &
\colhead{} &
\colhead{(eV)} &
\colhead{} &
\colhead{} }
\startdata
22386.90 & Fe~I & 5.033 & $-$0.490 & Solar \\
22399.16 & Fe~I & 5.099 & $-$1.275 & Solar \\
21785.60 & Si~I & 6.719 &  ~~0.295 & Solar \\
21825.63 & Si~I & 6.721 &  ~~0.030 & Solar \\
21788.63 & Ti~I & 1.749 & $-$0.884 & Arcturus \\
\enddata
\end{deluxetable}

\clearpage

\begin{deluxetable}{ccrrrrrr}
\tablecaption{Equivalent widths (m\AA) of program stars.}
\tablenum{9}
\tablewidth{0pc}
\tablehead{
\multicolumn{2}{c}{} &
\multicolumn{3}{c}{Palomar~6} &
\multicolumn{1}{c}{} &
\multicolumn{2}{c}{M71} \\
\cline{3-5}\cline{7-8}\\
\colhead{$\lambda$ (\AA)} &
\colhead{Id.} &
\colhead{A} &
\colhead{D} &
\colhead{G} &
\colhead{} &
\colhead{I-45} &
\colhead{I-46} }
\startdata
22386.90&Fe~I& 299    & 259    & 296 && 251 & 284 \\
22399.16&Fe~I& 130    & 142    & 120 && 152 & 160 \\
21785.60&Si~I&\nodata &\nodata & 226 && 251 & 174 \\
21825.63&Si~I&\nodata &\nodata & 194 && 238 & 174 \\
21788.63&Ti~I&\nodata &\nodata & 545 && 473 & 467 \\
\enddata
\end{deluxetable}

\clearpage

\begin{landscape}
\begin{deluxetable}{cccccccccc}
\tablecaption{Elemental abundances.}
\tablenum{10}
\tablewidth{0pc}
\tablehead{
\multicolumn{1}{c}{} &
\multicolumn{3}{c}{Palomar~6} &
\multicolumn{1}{c}{} &
\multicolumn{2}{c}{M71\tablenotemark{1}} &
\multicolumn{1}{c}{} &
\multicolumn{2}{c}{M71\tablenotemark{2}} \\
\cline{2-4}\cline{6-7}\cline{9-10}\\
\colhead{} &
\colhead{A} &
\colhead{D} &
\colhead{G} &
\colhead{} &
\colhead{I-45} &
\colhead{I-46} &
\colhead{} &
\colhead{I-45} &
\colhead{I-46} }
\startdata
$T_eff$ & 3718 & 3741 & 3667 && 4043 & 4019 && 4050 & 4000\\

$\log g$ & 0.3 & 0.3 & 0.2 && 1.0 & 1.0 && 0.8 & 0.8\\

$V_{\rm turb}$ & 2.00 & 2.00 & 2.00 && 2.00 & 2.15 && 2.00 & 2.15 \\

[Fe/H]& $-$1.0 $\pm$ 0.1 & $-$1.0 $\pm$ 0.1 & $-$1.0 $\pm$ 0.1 &&
$-$0.9 $\pm$ 0.1& $-$0.8 $\pm$ 0.1&&$-$0.78 $\pm$ 0.04&$-$0.79 $\pm$ 0.06 \\

[Si/Fe]&\nodata&\nodata&+0.4 $\pm$ 0.0 &&
+0.3 $\pm$ 0.1&$-$0.1 $\pm$ 0.1&&$+$0.38 $\pm$ 0.06&\nodata \\

[Ti/Fe]&\nodata  &\nodata & +0.5 &&
+0.6 & +0.4 & & +0.56 $\pm$ 0.06& +0.47 $\pm$ 0.07\\
\enddata
\tablenotetext{1}{This work.}
\tablenotetext{2}{Sneden et al.\ (1994).}

\end{deluxetable}
\end{landscape}

\clearpage

\begin{deluxetable}{cccc}
\tablecaption{Dependences of abundances on atmospheric parameters.}
\tablenum{11}
\tablewidth{0pc}
\tablehead{
\colhead{} &
\colhead{$\delta T_{\rm eff}$} &
\colhead{$\delta \log g$} &
\colhead{$\delta V_{\rm turb}$} \\
\colhead{} &
\colhead{(+200 K)} &
\colhead{(+0.2)} &
\colhead{(+0.5 km/s)} }
\startdata
[Fe/H] & +0.10 & +0.05 & $-$0.14 \\

[Si/Fe]& $-$0.12 & $-$0.05 & +0.02 \\

[Ti/Fe]& +0.29 & $-$0.03 & $-$0.33 \\
\enddata
\end{deluxetable}

\clearpage

\begin{deluxetable}{lcccr}
\tablecaption{$\cos \psi$ and $V_S$ for globular clusters with $R_{GC}$
$\leq$ 3 kpc.}
\tablenum{12}
\tablewidth{0pc}
\tablehead{
\colhead{ID} &
\colhead{[Fe/H]} &
\colhead{$R_{GC}$} &
\colhead{$\cos \psi$} &
\colhead{$V_S$} \\
\colhead{} &
\colhead{} &
\colhead{(kpc)} &
\colhead{} &
\colhead{(km sec$^{-1}$)}
}
\startdata
Terzan 3    & $-$0.73 &  2.4 &  $-$0.975 & $-$184.92 \\
NGC 6235    & $-$1.40 &  2.9 &  $-$0.085 & 94.77 \\
NGC 6256    & $-$0.70 &  2.1 &  $-$0.808 & $-$140.45 \\
NGC 6266 (M62)  & $-$1.29 &  1.7 &  $-$0.624 & $-$84.90 \\
NGC 6273 (M19)  & $-$1.68 &  1.6 &  $-$0.584 & 133.75 \\
NGC 6287    & $-$2.01 &  1.7 &  $+$0.052 & $-$275.91 \\
NGC 6293    & $-$1.99 &  1.4 &  $-$0.413 & $-$150.35 \\
NGC 6304    & $-$0.59 &  2.1 &  $-$0.291 & $-$113.33 \\
NGC 6325    & $-$1.17 &  2.0 &  $+$0.089 & 45.09 \\
NGC 6333 (M9)   & $-$1.72 &  1.7 &  $+$0.975 & 262.87 \\
NGC 6342    & $-$0.65 &  1.7 &  $+$0.792 & 147.42 \\
NGC 6355    & $-$1.50 &  1.0 &  $-$0.070 & $-$167.61 \\
Terzan 2    & $-$0.40 &  0.9 &  $-$0.586 & 104.59 \\
Terzan 4    & $-$1.60 &  1.4 &  $-$0.445 & $-$55.27 \\
HP 1        & $-$1.50 &  0.8 &  $-$0.516 & 53.20 \\
Liller 1    & $-$0.30 &  2.6 &  $-$0.273 & 41.21 \\
Terzan 1    & $-$0.35 &  1.8 &  $-$0.186 & 35.57 \\
Ton 2       & $-$0.50 &  1.4 &  $-$0.988 & $-$212.01 \\
NGC 6401    & $-$1.12 &  0.8 &  $+$0.843 & $-$39.99 \\
Palomar 6   & $-$1.08 &  0.9 &  $+$0.343 & 200.22 \\
Terzan 5    & $-$0.28 &  0.7 &  $+$0.809 & $-$67.79 \\
NGC 6440    & $-$0.34 &  1.3 &  $+$0.919 & $-$36.37 \\
Terzan 6    & $-$0.50 &  1.6 &  $-$0.132 & 130.21 \\
Terzan 9    & $-$1.00 &  0.6 &  $+$0.866 & 83.81 \\
NGC 6522    & $-$1.44 &  0.6 &  $+$0.547 & $-$7.09 \\
NGC 6528    & $+$0.07 &  1.3 &  $+$0.146 & 199.37 \\
NGC 6541    & $-$1.76 &  2.2 &  $-$0.816 & $-$202.01 \\
NGC 6553    & $-$0.16 &  2.5 &  $+$0.294 & 24.90 \\
NGC 6558    & $-$1.44 &  1.0 &  $+$0.043 & $-$186.84 \\
NGC 6569    & $-$0.86 &  1.2 &  $+$0.103 & $-$16.67 \\
NGC 6624    & $-$0.42 &  1.2 &  $+$0.976 & 74.51 \\
NGC 6626 (M28)  & $-$1.45 &  2.6 &  $+$0.432 & 58.22 \\
NGC 6638    & $-$0.99 &  1.6 &  $+$0.929 & 59.40 \\
NGC 6637 (M69)  & $-$0.71 &  1.6 &  $+$0.451 & 55.80 \\
NGC 6642    & $-$1.35 &  1.6 &  $+$0.980 & $-$8.05 \\
NGC 6652    & $-$0.96 &  2.4 &  $+$0.146 & $-$96.84 \\
NGC 6681 (M70)  & $-$1.51 &  2.1 &  $+$0.436 & 240.28 \\
NGC 6717    & $-$1.29 &  2.3 &  $+$0.941 & 82.76 \\
NGC 6723    & $-$1.12 &  2.6 &  $+$0.023 & $-$86.44 \\
\enddata
\end{deluxetable}

\clearpage

\begin{figure}
\epsscale{1}
\figurenum{1}
\plotone{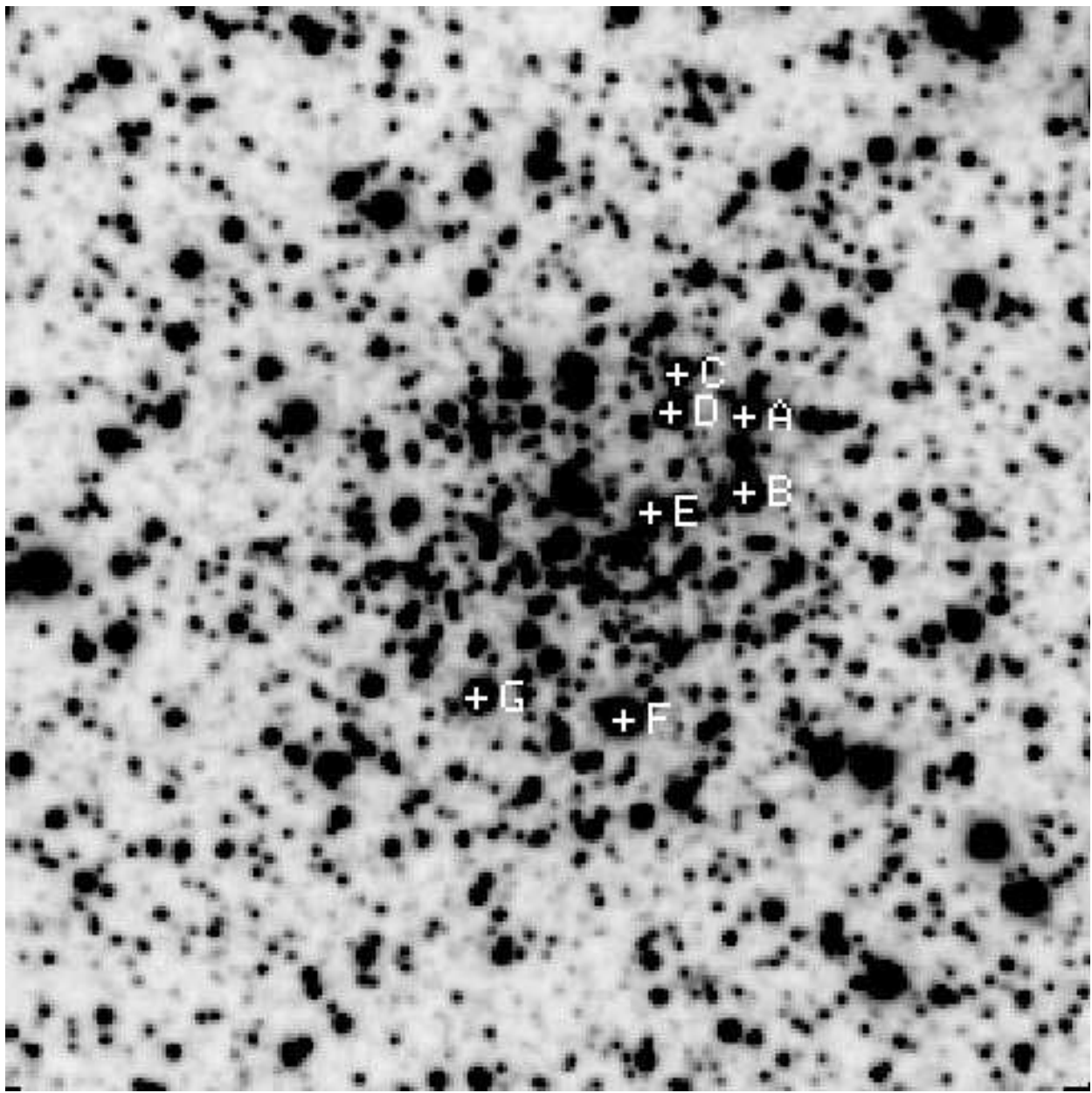}
\caption
{The finding chart of Palomar~6 spectroscopy program stars in $K$ passband.
The field of view is 2.56 $\times$ 2.56 arcmin.
North is at the top and east is to the left.}
\end{figure}

\clearpage
\begin{figure}
\epsscale{1}
\figurenum{2}
\plotone{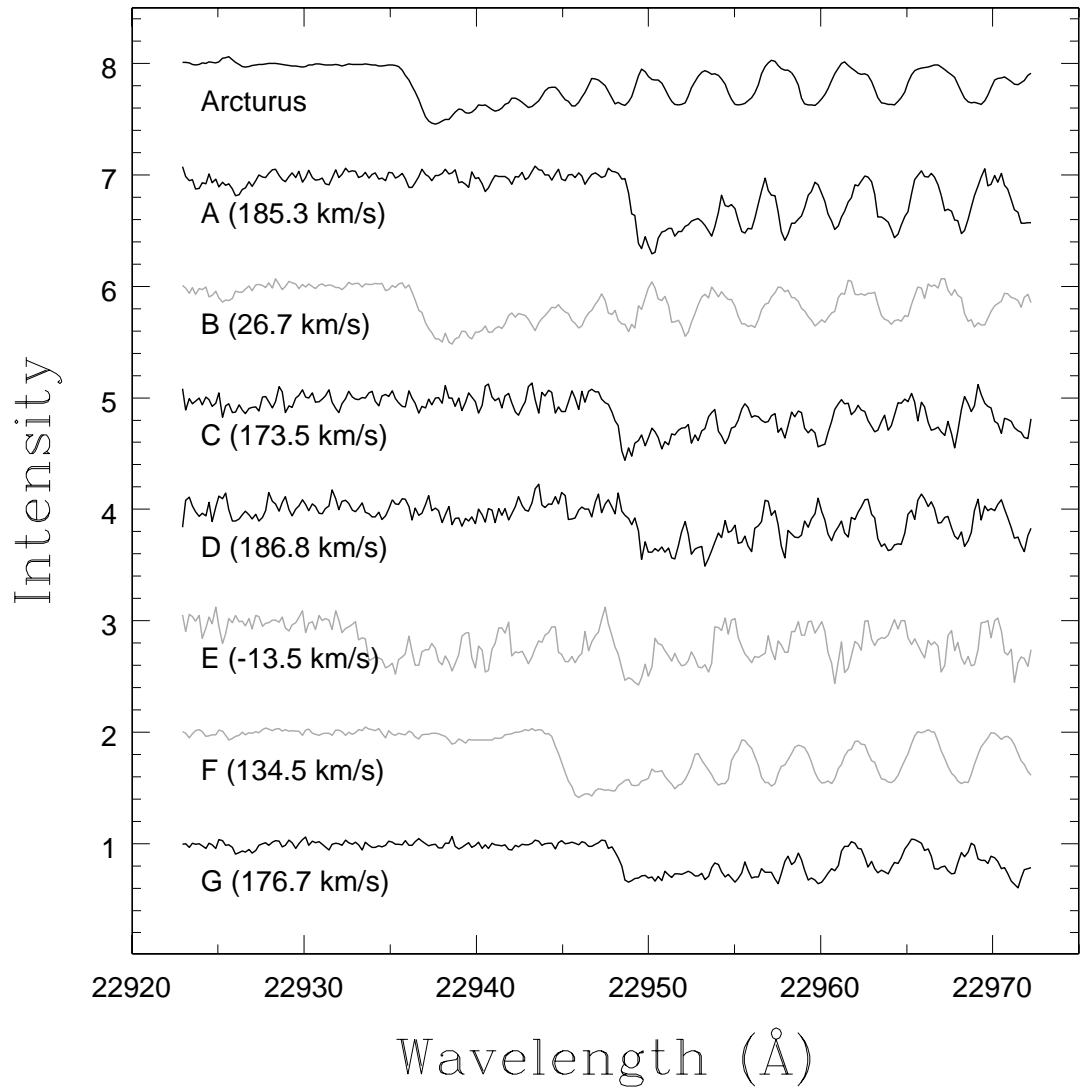}
\caption
{Heliocentric radial velocity measurements of Palomar~6 program stars.
The Palomar 6 member stars are presented by black lines and
nonmember stars by gray lines.}
\end{figure}

\clearpage
\begin{figure}
\epsscale{1}
\figurenum{3}
\plotone{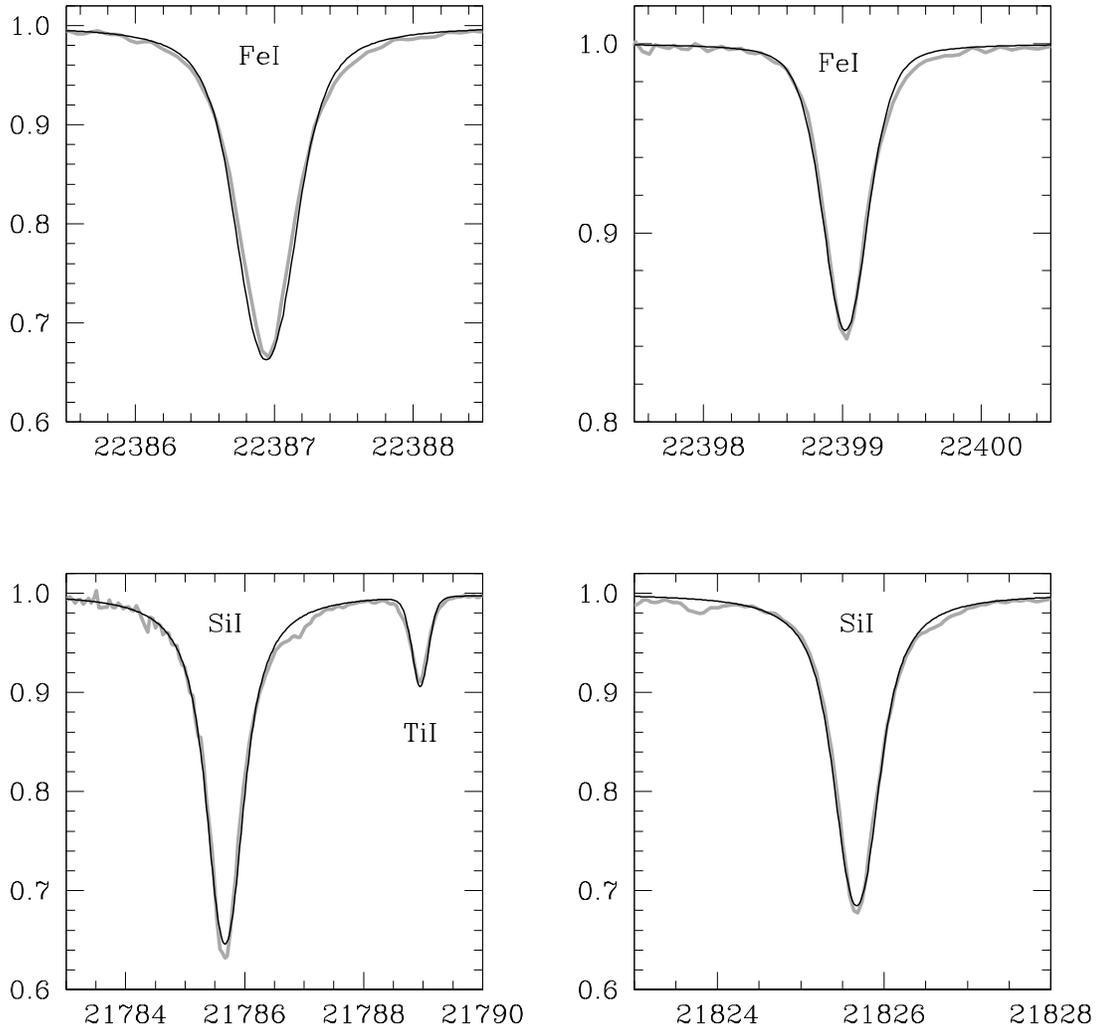}
\caption
{Comparisons of the synthetic solar spectra with the observed solar spectra.
Note that the observed solar spectra shows line asymmetries.}
\end{figure}

\clearpage
\begin{figure}
\epsscale{1}
\figurenum{4}
\plotone{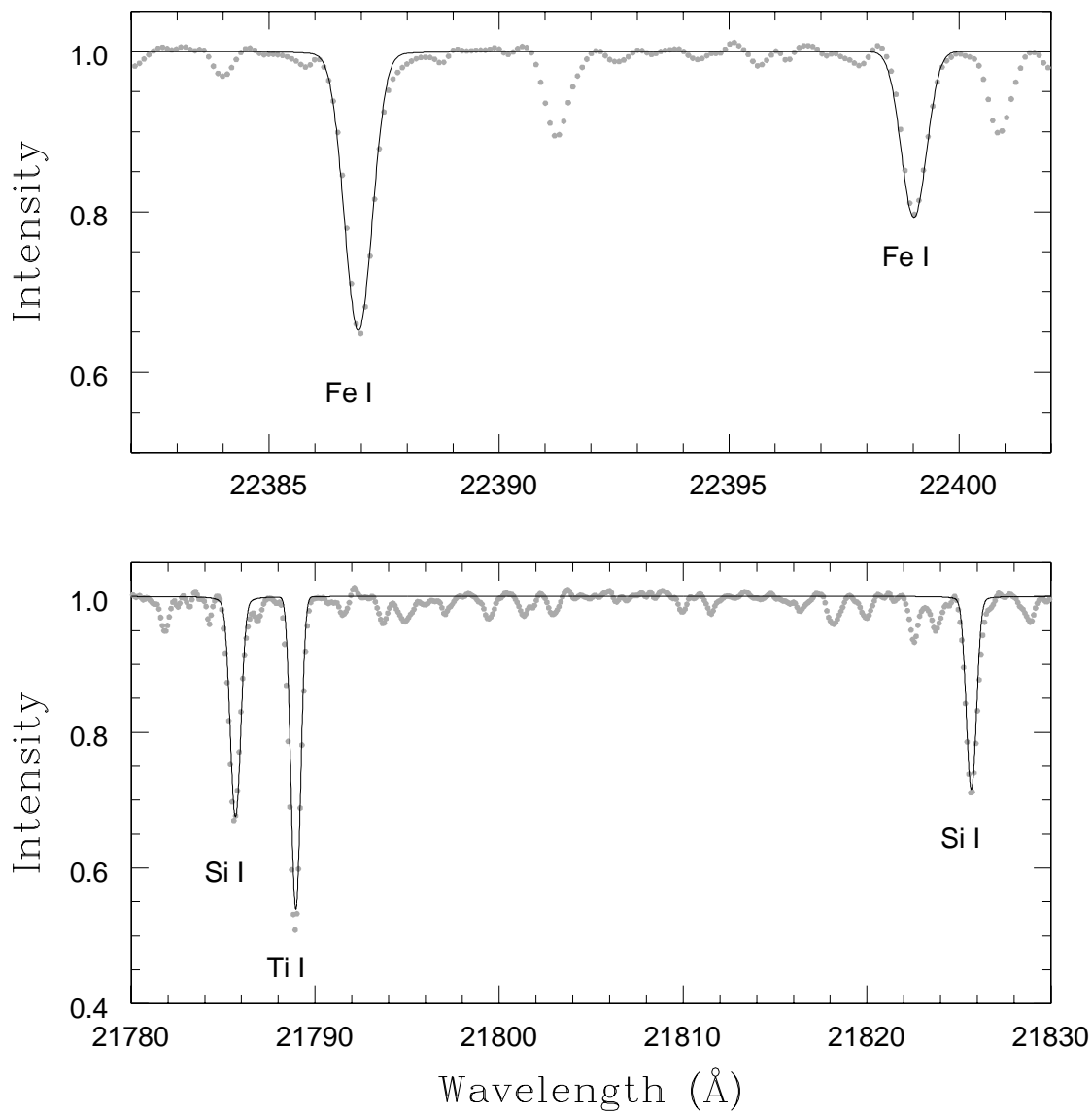}
\caption
{A comparison between the Arcturus spectrum (Hinkle et al.\ 1995)
and synthetic spectrum with solar $\log gf$ values
for [Fe/H] = $-$0.47, [Si/Fe] = +0.14 and [Ti/Fe] = +0.27
(Sneden et al.\ 1994).
Note that the solar $gf$ value for Ti~I fails to reproduce the observed
absorption line of Arcturus.}
\end{figure}

\clearpage
\begin{figure}
\epsscale{1}
\figurenum{5}
\plotone{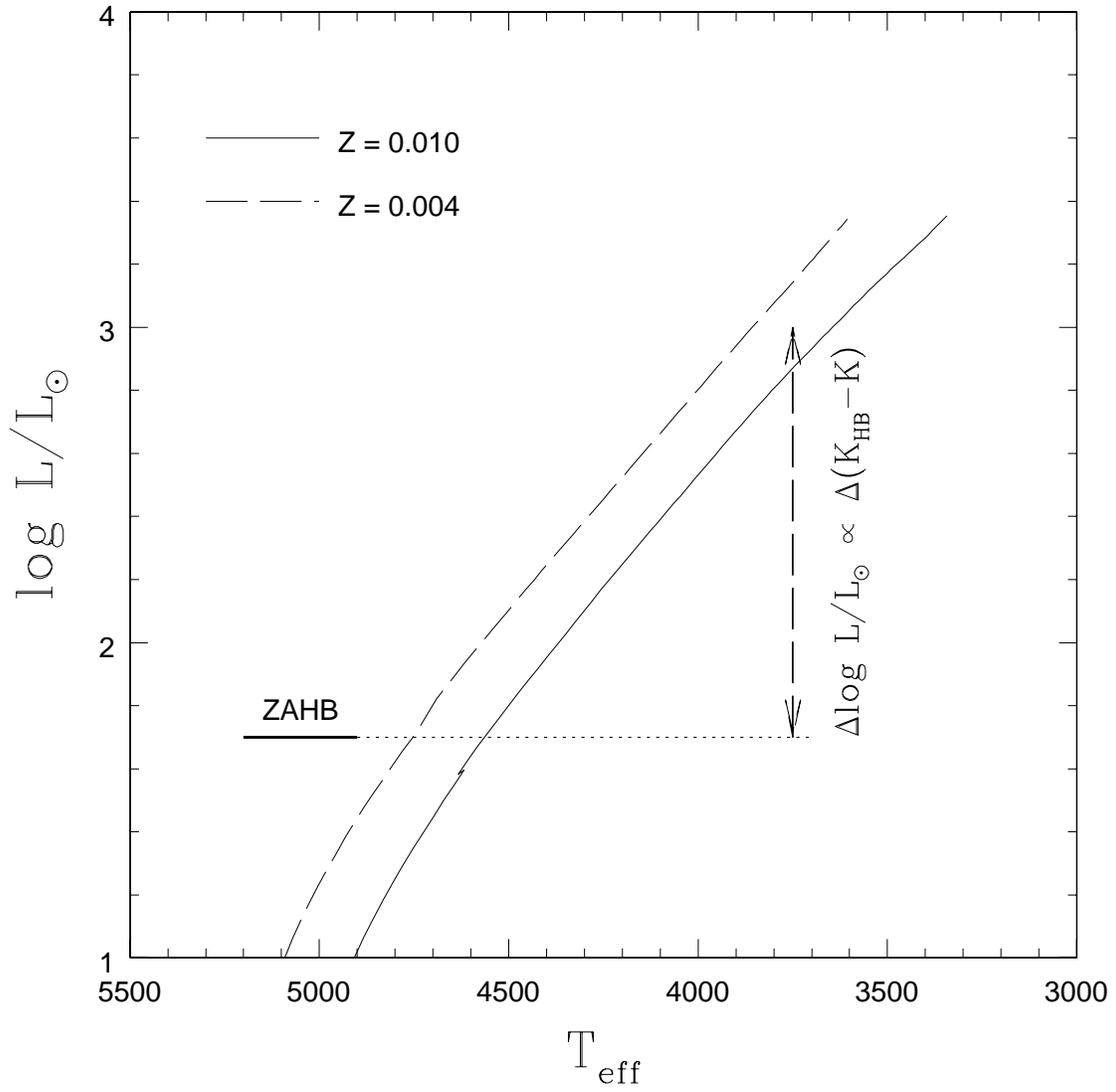}
\caption
{A schematic plot of $T_{\rm eff}$ versus $\log L/L_\odot$
for metal-rich RGB and HB sequences.}
\end{figure}

\clearpage
\begin{figure}
\epsscale{1}
\figurenum{6}
\plotone{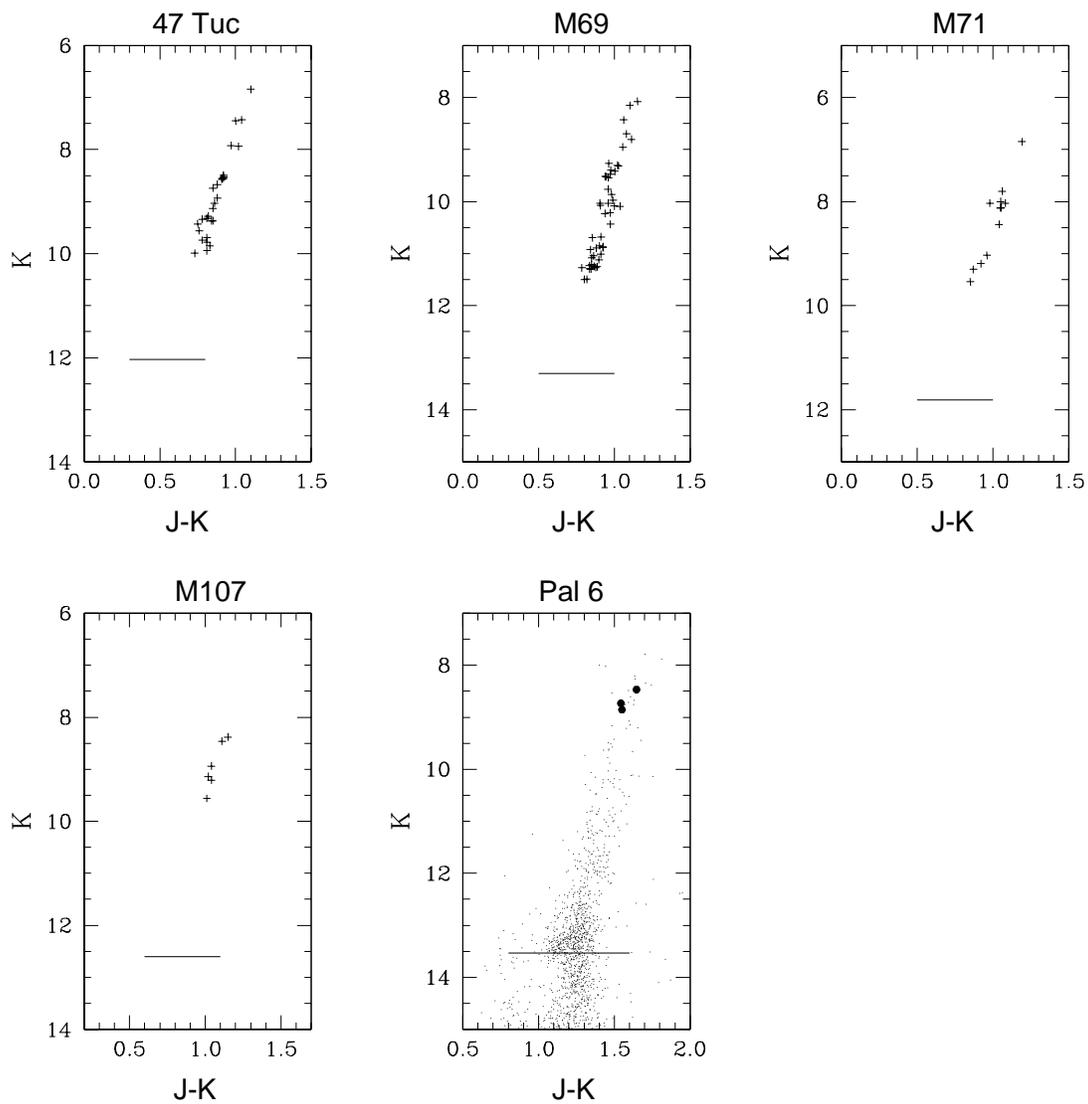}
\caption
{$JK$ CMDs of metal-rich globular clusters.
The $K_{(RGB,RHB)}$ is shown by the horizontal solid lines,
determined by Kuchinski et al.\ (1995).
The black dots represent our program RGB stars in Palomar~6.}
\end{figure}

\clearpage
\begin{figure}
\epsscale{1}
\figurenum{7}
\plotone{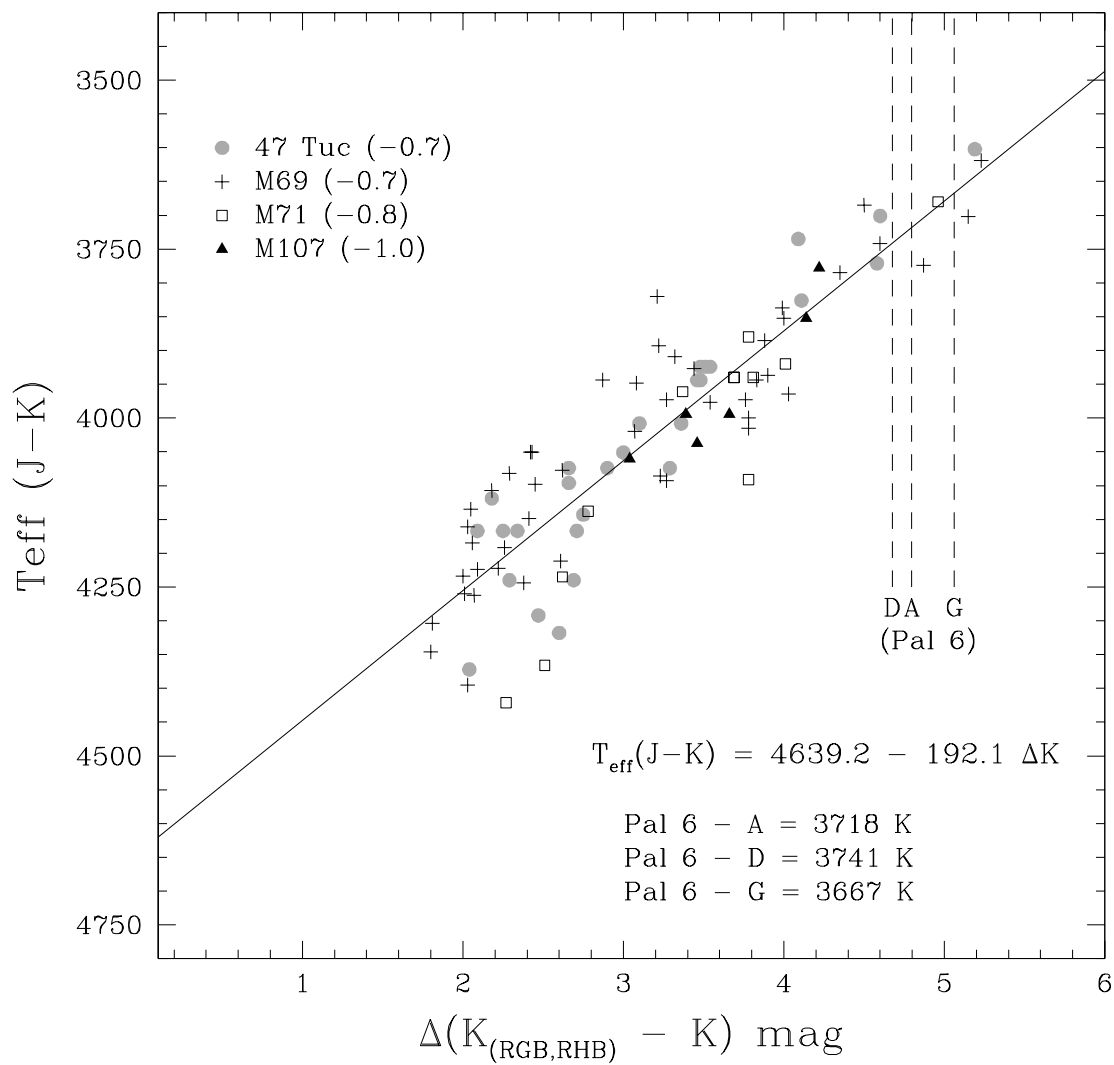}
\caption
{$T_{\rm eff}$ versus $\Delta(K_{(RGB,RHB)}-K)$ mag of RGB stars
in metal-rich clusters. The vertical dashed lines represent
$\Delta(K_{(RGB,RHB)}-K)$ mag of our program stars in Palomar~6.}
\end{figure}

\clearpage
\begin{figure}
\epsscale{1}
\figurenum{8}
\plotone{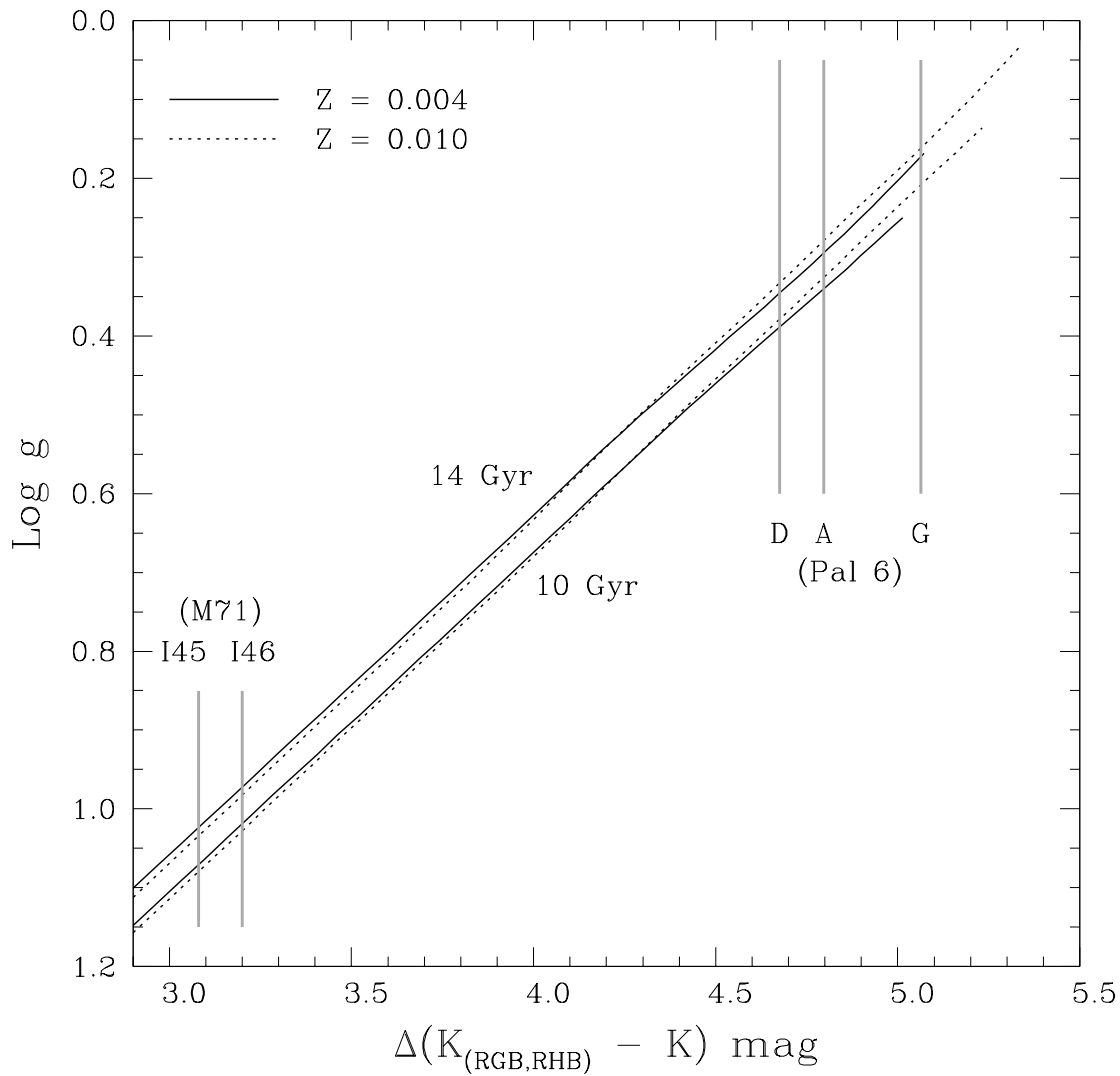}
\caption
{$\Delta (K_{(RGB,RHB)}-K)$ versus $\log g$
for the model isochrones with Z = 0.004 ([Fe/H] = $-$0.9),
Z = 0.010 ([Fe/H] = $-$0.5) and
[$\alpha$/Fe] = +0.3 for 10, 14 Gyr (Kim et al. 2002).
We also show the locations of our program stars in the plot.}
\end{figure}

\clearpage
\begin{figure}
\epsscale{1}
\figurenum{9}
\plotone{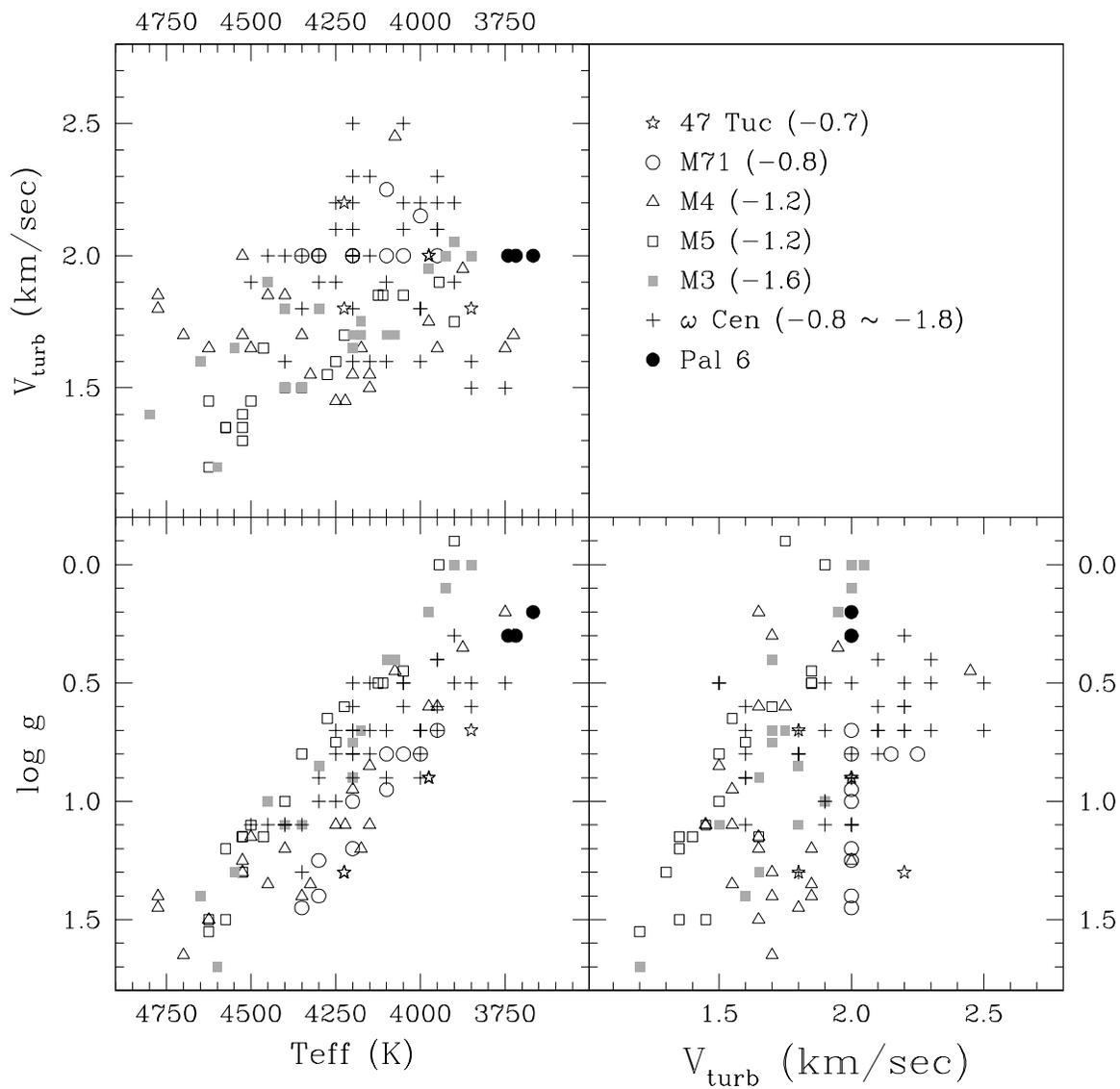}
\caption
{Comparisons of $\log g$ versus $T_{\rm eff}$ versus $V_{\rm turb}$
of RGB stars in globular clusters studied in optical.}
\end{figure}

\clearpage
\begin{figure}
\epsscale{1}
\figurenum{10}
\plotone{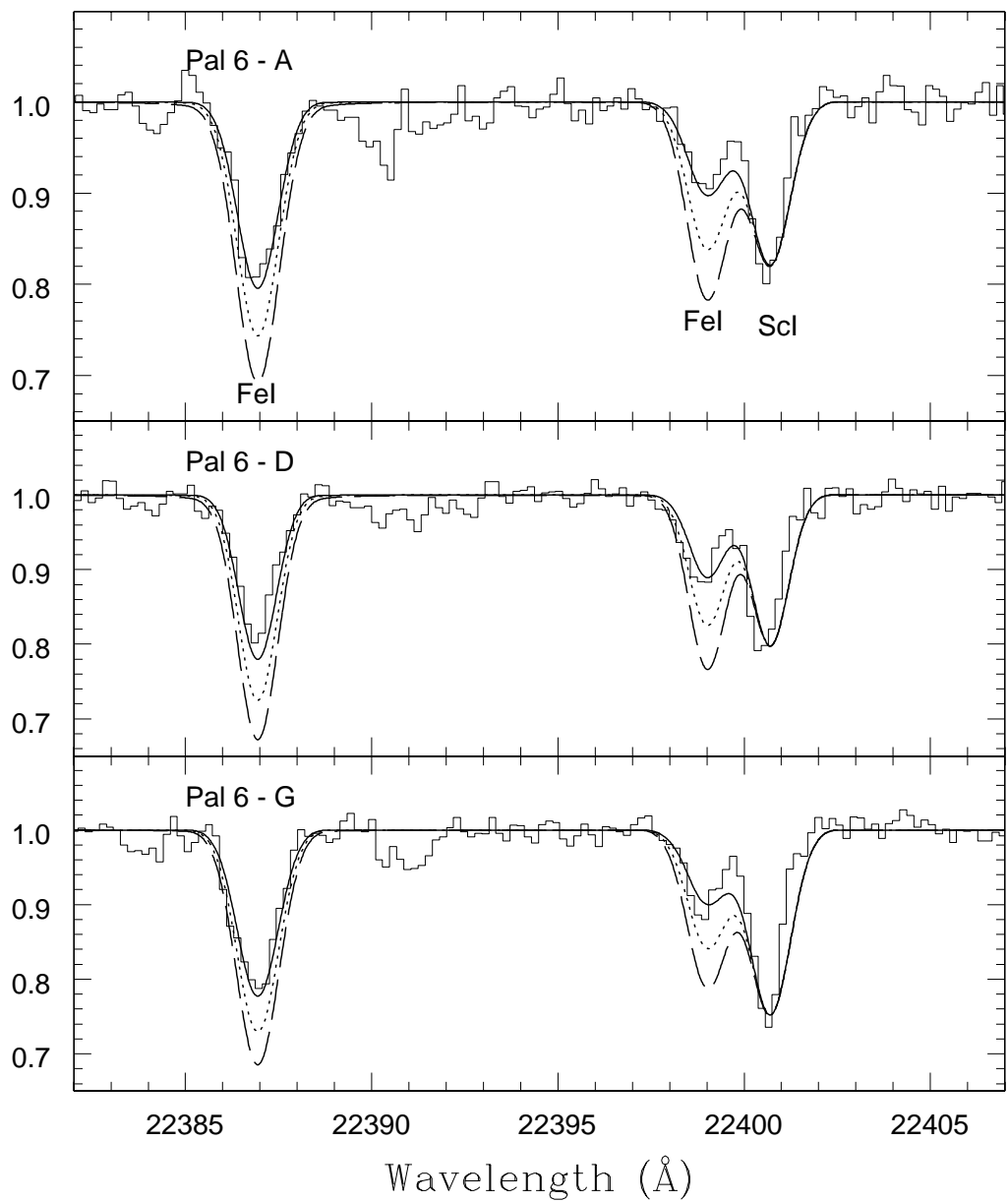}
\caption
{Comparisons of Fe I lines between observed and synthetic spectra.
Solid lines are for [Fe/H] = $-$1.0, dotted lines are for [Fe/H] = $-$0.5,
and dashed lines are for [Fe/H] = 0.0.
In the Figure, we adopt an ad hoc $gf$ value for the Sc~I line.}
\end{figure}

\clearpage
\begin{figure}
\epsscale{1}
\figurenum{11}
\plotone{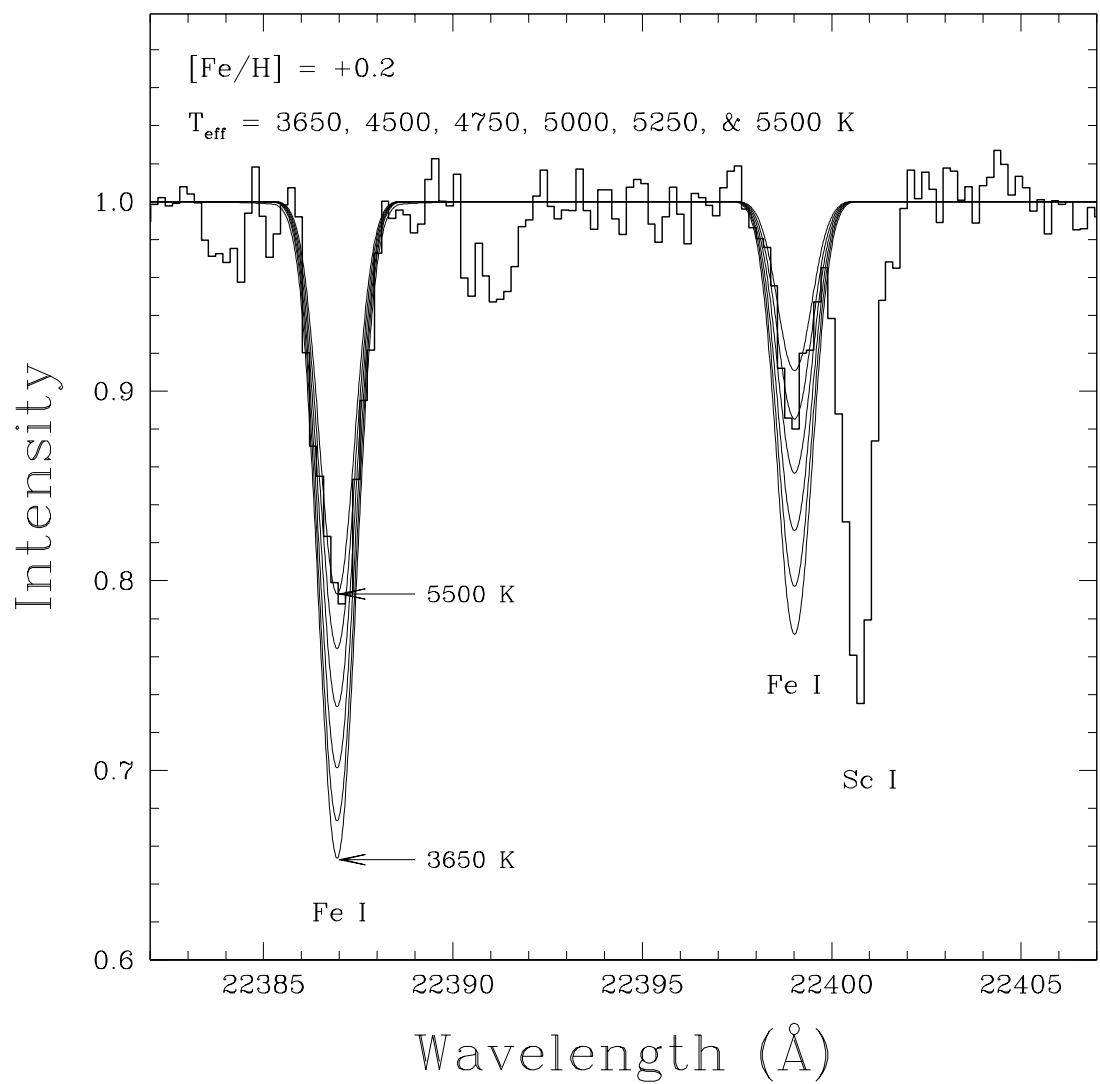}
\caption
{Synthetic spectra for Palomar~6-G using the model atmospheres with
[Fe/H] = +0.20.
Synthetic spectra do not reproduce the observed spectrum with
reasonable adjustment ($\pm$ 100 K) of the effective temperature.}
\end{figure}

\clearpage
\begin{figure}
\epsscale{1}
\figurenum{12}
\plotone{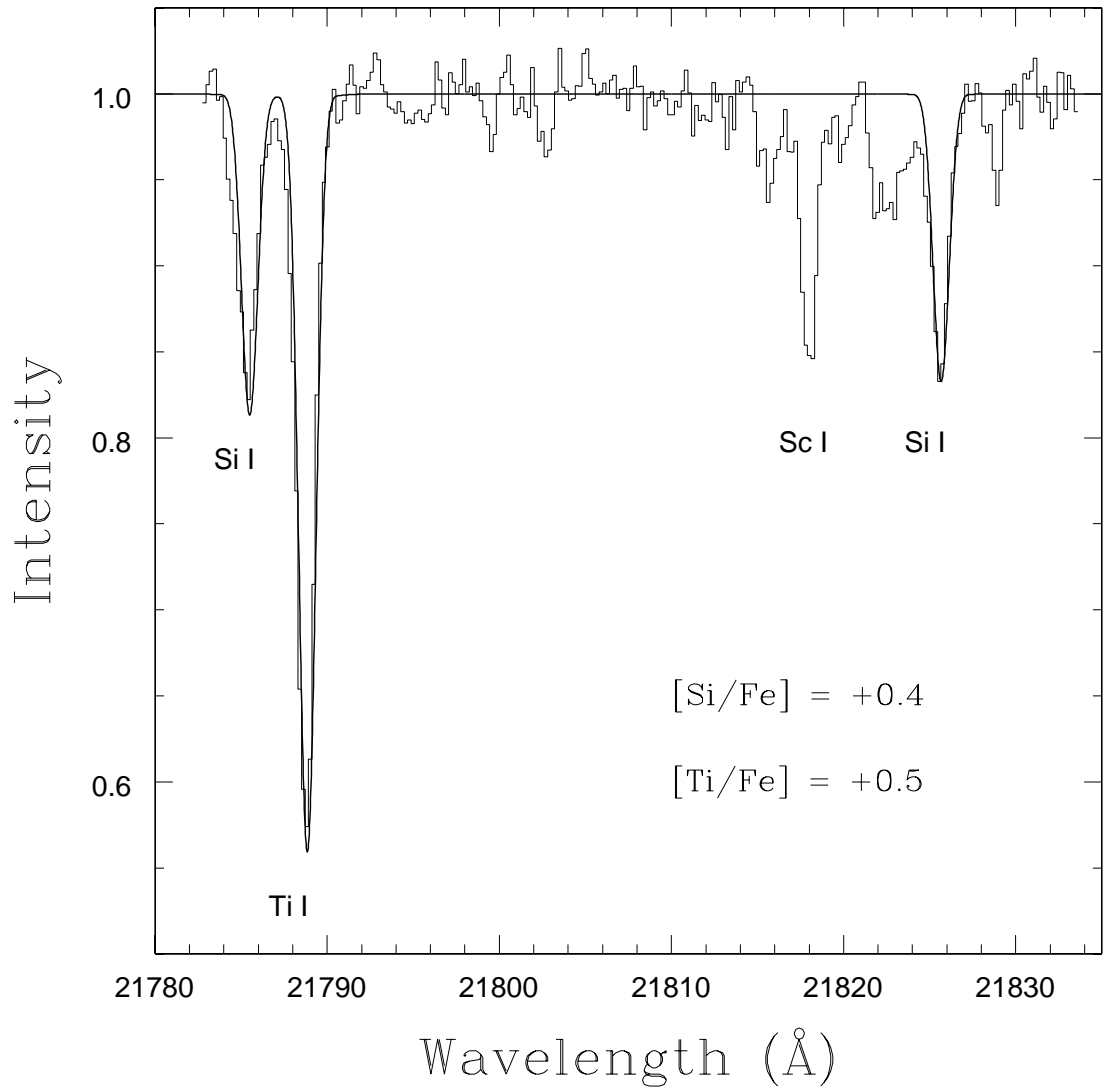}
\caption
{Comparisons of Si~I and Ti~I lines between observed and synthetic spectra.}
\end{figure}

\begin{figure}
\epsscale{1}
\figurenum{13}
\plotone{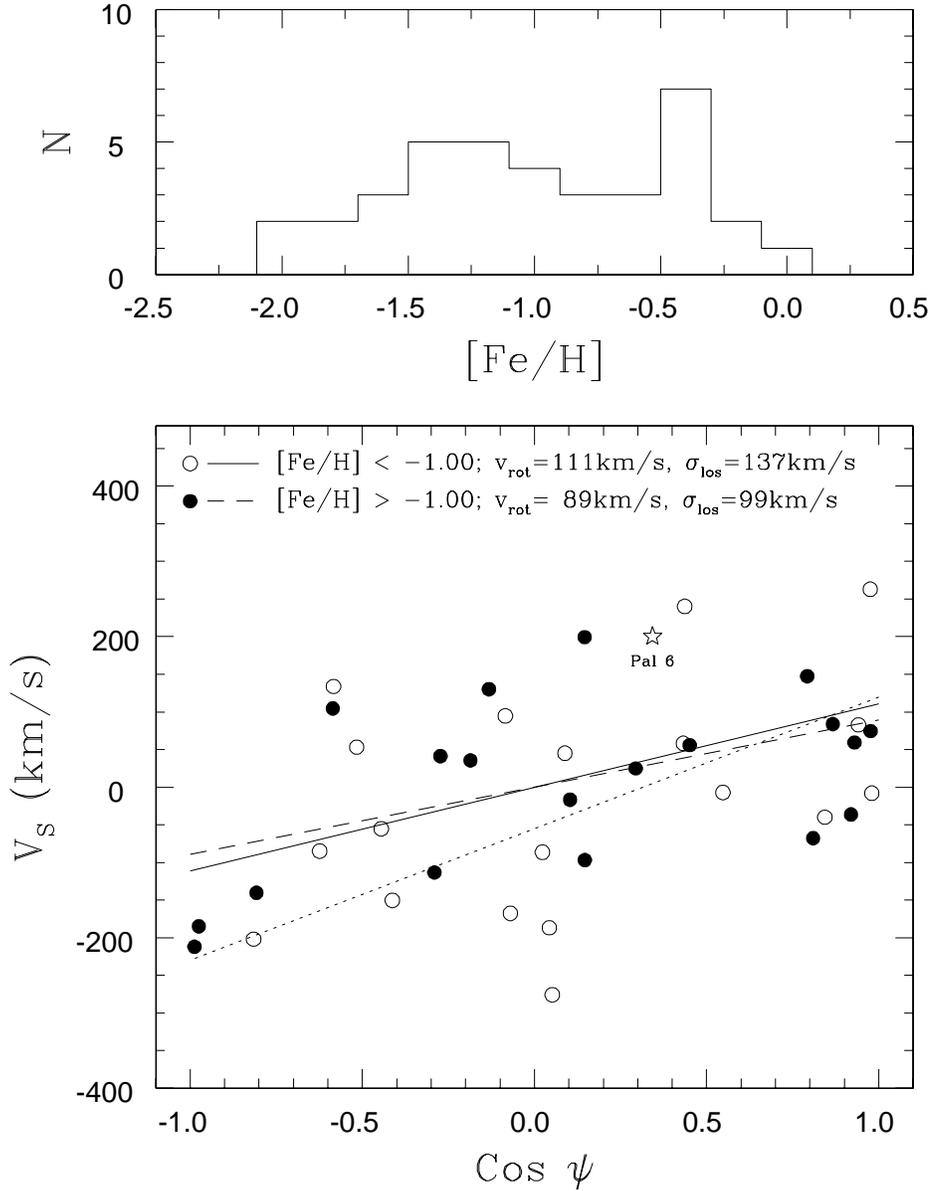}
\caption
{The metallicity distribution and the kinematical properties
of globular clusters with $R_{GC} \leq $ 3 kpc.
The metallicity distribution shows two peaks at [Fe/H] $\approx$
$-$1.3 and $-$0.4.
In the bottom panel, V$_S$ represents the radial velocity observed
at the Sun's position by an observer at rest with respect to the Galactic
center and $\psi$ the angle between the line of sight and the direction of
Galactic rotation at the cluster.
The mean rotational velocity is the slope of the straight line.
The open circles represent the low metallicity clusters and the filled
circles the high metallicity clusters.
The solid line represents the metal-poor clusters and the dashed line
the metal-rich clusters.
The rotational velocity solution of the metal-rich
clusters with the lower mass bar-like kinematics
is also presented by dotted line (Burkert \& Smith 1997).}
\end{figure}

\clearpage
\begin{figure}
\epsscale{1}
\figurenum{14}
\plotone{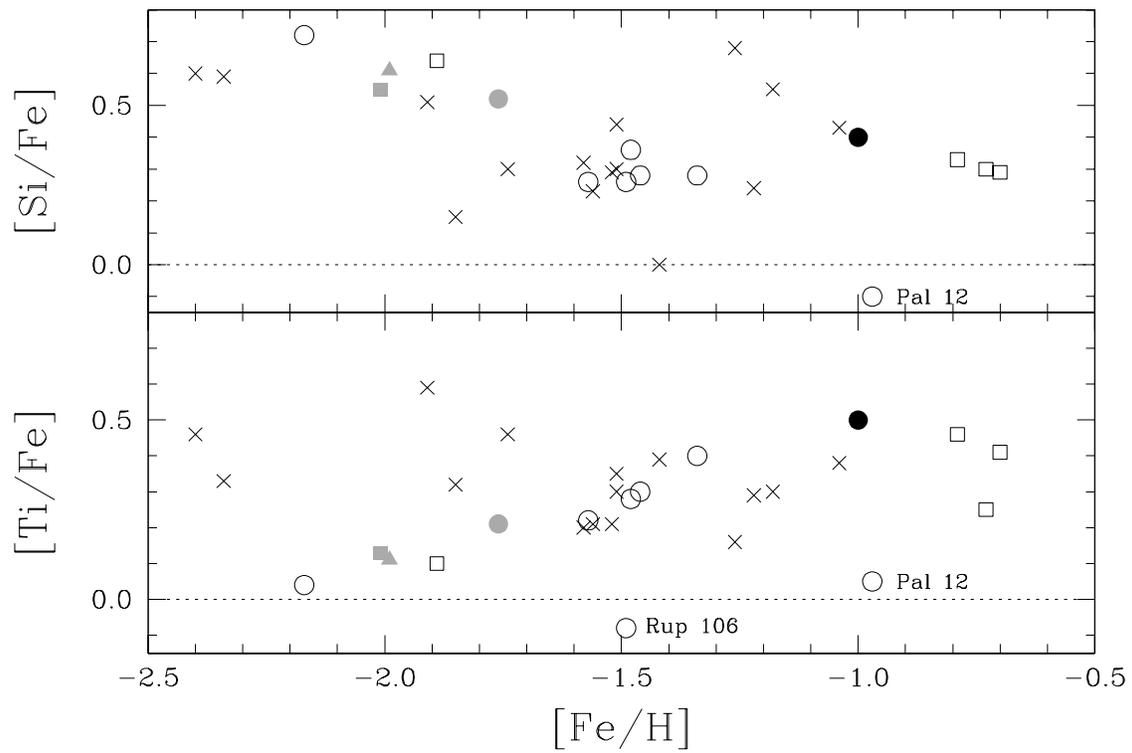}
\caption
{Silicon and titanium abundances for globular clusters.
Crosses represent ``Old Halo" clusters, open circles ``Younger Halo" clusters,
open squares ``Disk" cluster.
NGC~6287 is represented by gray squares, NGC~6293 by gray triangles,
NGC~6541 by gray circles, and
Palomar~6 is represented by black circles.}
\end{figure}

\clearpage
\begin{figure}
\epsscale{1}
\figurenum{15}
\plotone{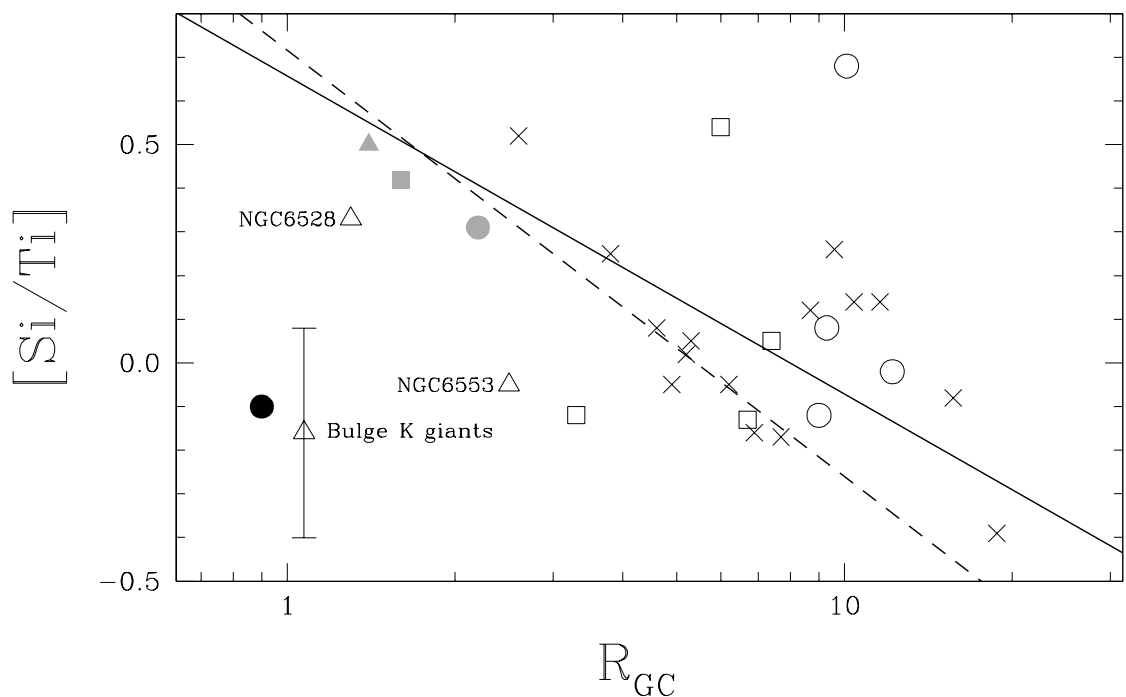}
\caption
{[Si/Ti] ratio as a function of $R_{GC}$.
Crosses represent ``Old Halo" clusters, open circles ``Younger Halo" clusters,
open squares ``Disk" cluster.
NGC~6287 is represented by a gray square, NGC~6293 by a gray triangle,
NGC~6541 by a gray circle, NGC~6528 and NGC~6553 by open triangles,
and Palomar~6 is represented by black circles.
The solid line for the bisector linear fit to the old halo clusters
(18 clusters) and the dashed line represents the bisector linear fit
to the old halo clusters with $R_{GC}$ $\leq$ 8 kpc (12 clusters).
The open triangle with an error bar indicates the [Si/Ti] ratio of bulge K
giants by McWilliam \& Rich (1994) at an arbitrary Galactocentric distance.}
\end{figure}

\end{document}